\documentclass[10pt,journal,compsoc]{IEEEtran}

\usepackage{url}
\usepackage{tikz}
\usepackage{times}
\usepackage{grffile}
\usepackage{siunitx}
\usepackage{pgfplots}
\usepackage{latexsym}
\usepackage{graphicx}
\usepackage{amsfonts}
\usepackage{booktabs}
\usepackage{subfigure}
\usepackage{amsmath,bm} 
\usepackage{environ}
\usepgfplotslibrary{units}

\pgfplotsset{width=6cm, compat=1.3}

\ifCLASSOPTIONcompsoc
  \usepackage[nocompress]{cite}
\else
  \usepackage{cite}
\fi

\pgfplotsset{compat=1.14}

\begin{document}
\title{CED: Credible Early Detection\\ of Social Media Rumors}

\author{Changhe Song, Cunchao Tu, Cheng Yang, Zhiyuan Liu, Maosong Sun
\IEEEcompsocitemizethanks{\IEEEcompsocthanksitem Cunchao Tu, Cheng Yang, Zhiyuan Liu (corresponding author), and Maosong Sun are with the Department of Computer Science and Technology, Tsinghua University, Beijing 100084, China.\protect\\
E-mail: tucunchao@gmail.com, cheng-ya14@mails.tsinghua.edu.cn, \{liuzy, sms\}mail.tsinghua.edu.cn
\IEEEcompsocthanksitem Changhe Song is with the Department of Electronic Engineering, Tsinghua University, Beijing 100084, China.\protect\\
E-mail: sch14@mails.tsinghua.edu.cn}
\thanks{Manuscript received April 19, 2005; revised August 26, 2015.}}

\markboth{Journal of \LaTeX\ Class Files,~Vol.~14, No.~8, August~2015}%
{Shell \MakeLowercase{\textit{et al.}}: Bare Demo of IEEEtran.cls for Computer Society Journals}

\IEEEtitleabstractindextext{
\begin{abstract}
Rumors spread dramatically fast through online social media services, and people are exploring methods to detect rumors automatically. Existing methods typically learn semantic representations of all reposts to a rumor candidate for prediction. However, it is crucial to efficiently detect rumors as early as possible before they cause severe social disruption, which has not been well addressed by previous works. In this paper, we present a novel early rumor detection model, Credible Early Detection (CED). By regarding all reposts to a rumor candidate as a sequence, the proposed model will seek an early point-in-time for making a credible prediction. We conduct experiments on three real-world datasets, and the results demonstrate that our proposed model can remarkably reduce the time span for prediction by more than $85\%$, with better accuracy performance than all state-of-the-art baselines. 
\end{abstract}

}

\maketitle
\IEEEdisplaynontitleabstractindextext
\IEEEpeerreviewmaketitle

\setkeys{Gin}{draft=false}

\IEEEraisesectionheading{\section{Introduction}\label{sec:introduction}}

\IEEEPARstart{R}{umor} is an important phenomenon in social science and has witnessed many interests of researchers in social psychology field for many decades~\cite{allport1947psychology,kapferer1987rumeurs}. According to the explanation of wikipedia\footnote{https://en.wikipedia.org/wiki/Rumor} and sociologists ~\cite{peterson1951rumor}, a rumor usually involves some concerned public statements whose integrity cannot be quickly or ever verified.

With the rapid growth of large-scale social media platforms, such as Facebook, Twitter, and Sina Weibo, rumor is becoming a more and more serious social problem than ever before. Due to the convenience of accessing information on these social media platforms, rumors can spread explosively within a short time before contradicted or detected. The widespread of rumors in social media severely prevents people achieving reliable information and may result in enormous economic loss or public panic when some emergencies happen.

How to detect rumors at an early stage of spread is extremely critical to prevent these damages. However, it is complicated for ordinary people to distinguish rumors from massive amounts of online information, due to the limitation of professional knowledge, time or space ~\cite{liu2015rumor}. Therefore, many news organizations and social media service providers pay great efforts to construct rumor reporting and collecting websites (e.g.,snopes\footnote{http://www.snopes.com/} and factcheck\footnote{http://www.factcheck.org/}) or platforms (e.g., Sina Rumor Reporting Center). Nevertheless, such practice needs a lot of human efforts to collect and verify rumors, and also faces issues of coverage and time delay.

\begin{figure}[!t]
	\centering
	\includegraphics[width=\columnwidth]{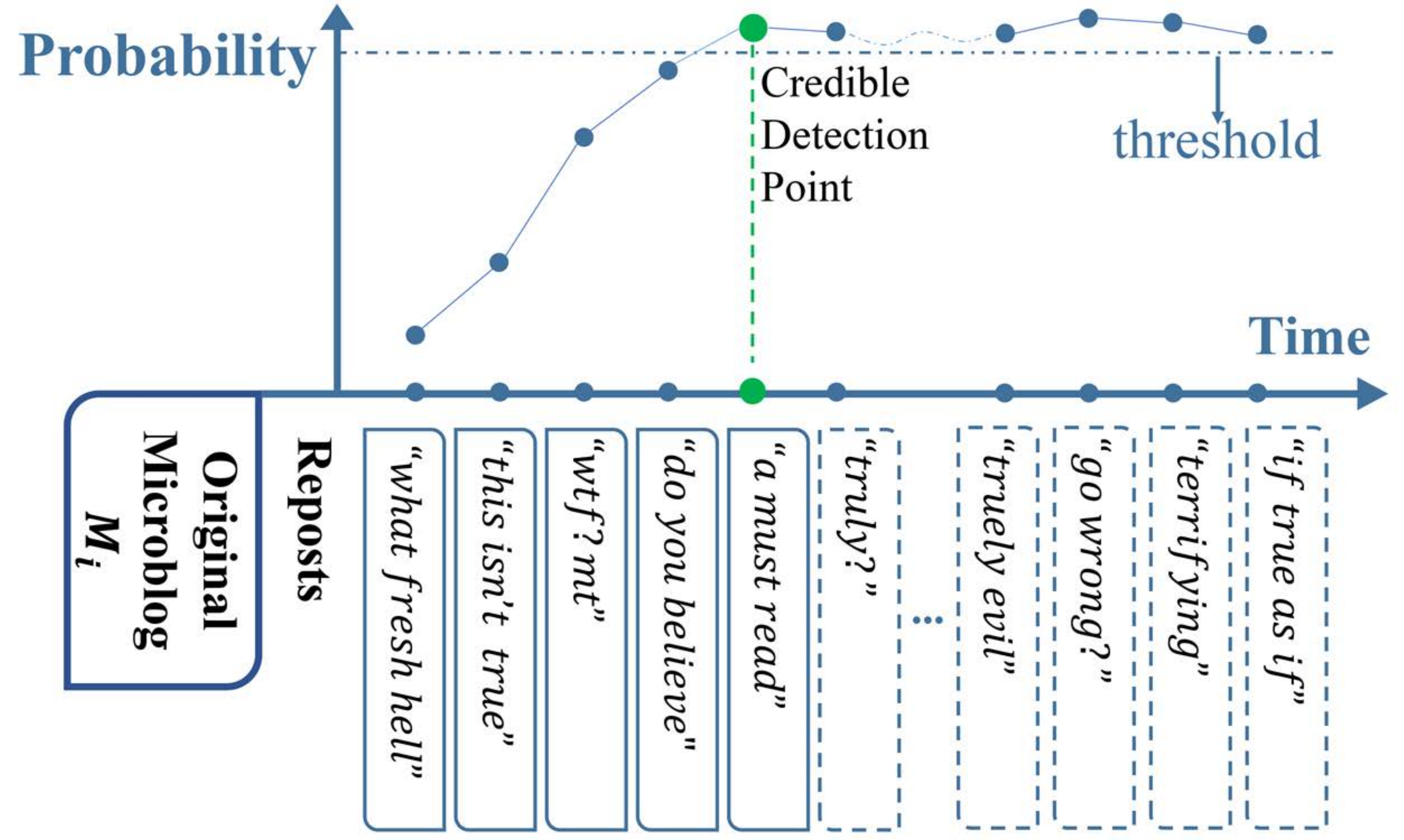}
	\caption{An example of early rumor detection with Credible Detection Point.}
	\label{fig:early-detection}
\end{figure}

To address these issues, researchers propose to detect online rumors automatically using machine learning techniques. Most existing models treat rumor detection as a binary classification task and tend to design effective features from various information sources, including text content, publisher's profiles and propagation patterns~\cite{castillo2011information,yang2012automatic,kwon2013prominent,liu2015real,ma2015detect,wu2015false}. However, these feature engineering-based methods are biased and time-consuming, and can not ensure the flexibility and generalization when applied to other rumor detection scenarios.

Compared with feature engineering methods, the effectiveness of deep neural networks on automatic feature learning has been verified in many NLP tasks, such as parsing, text classification, machine translation and question answering. Motivated by the successful utilization of deep neural networks, Ma et al. ~\cite{ma2016detecting} employed recurrent neural network (RNN) to learn a dynamic temporal representation for each microblog based on its reposts over time and make the prediction according to the representation of the entire repost sequence. It is the first attempt to introduce deep neural networks into repost-based rumor detection and achieves considerable performance on real-world datasets. Besides, Yu et al.~\cite{yu2017convolutional} employed paragraph vector~\cite{le2014distributed} and convolutional neural networks on the repost sequence to detect rumors.

However, to achieve early rumor detection, the model needs to make a reliable prediction as early as possible before its widespread. It is worth pointing out that existing neural network based rumor detection methods can only detect rumors by considering the entire or fixed proportions of repost information, which is not effective in practice. In this article, we propose a novel model to implement online real-time detection of social media rumors.

As shown in Fig.~\ref{fig:early-detection}, we present the temporal repost sequence of a specific rumor and the corresponding prediction probability curve. For this example, we can make a credible prediction at an early time stamp, as there appears quite a lot doubts and refutation to the original microblog. Based on this observation, we introduce ``Credible Detection Point'' and propose a novel early rumor detection model, Credible Early Detection (CED). Specifically, CED learns to determine the ``Credible Detection Point'' for each repost sequence during the training stage and ensure the credibility of the prediction result at this time point, i.e., there is no plot reversal after credible detection point. In this way, we can also make reliable detection without using the information after the ``Credible Detection Point'' in real applications, so that to detect rumors as soon as possible in social media platforms. 

To verify the effectiveness of our model, we conduct rumor detection experiments on three representative real-world datasets, and experimental results demonstrate that our proposed model can remarkably reduce the time span for prediction by more than $85\%$, even with much better detection accuracy.

We make the following noteworthy contributions:

(1) To solve the problem that current rumor detection methods can only detect rumors using the entire or fixed proportions of repost information, we proposed the concept of ``Credible Detection Point'', making it possible to dynamically obtain the minimum amount of repost information needed to make a reliable detection during detection process for each microblog.  

(2) Based on ``Credible Detection Point'', we propose a novel repost-based early rumor detection model, Credible Early Detection (CED). By regarding all reposts of a microblog as a sequence, CED will seek a microblog-specific ``Credible Detection Point'' for each microblog to make a credible prediction as early as possible.


(3) We conduct a series of experiments on three real-world datasets. Experimental results demonstrate that our model can remarkably reduce the detection time by more than $85\%$, with even better detection accuracy than all state-of-the-art baselines. Moreover, more than $60\%$ percent microblogs can be detected with less than $10\%$ repost information. This performance makes real-time monitoring of rumors on social media possible.

(4) We expand the relevant Weibo rumors dataset ~\cite{ma2016detecting} about twice the original size. We believe that the expanded dataset can better benefit further researches on rumor detection.


We will release all source codes and datasets of this work on Github~\footnote{https://github.com/} for further research explorations.

\section{Related Work}
The study of rumors can be traced back to 1940s~\cite{allport1947psychology}. With the rise of social networking media in recent decades, more and more methods of rumor detection have received a lot of attention. Most rumors detecting methods have been included in the comprehensive survey of Kumar et al~\cite{kumar2018false}.

Rumor detection in social media aims to distinguish whether a message posted on social media platforms is a rumor or not according to its relevant information, such as text content, comments, repost patterns, publisher's profiles and so on. Based on the types of information and methods used, we divide the existing rumor detection methods into three categories: (1) Traditional classification methods using artificially designed feature; (2) Deep neural networks related methods; (3) Propagation mode based methods. 


\subsection{Traditional Classification Methods}
Most traditional proposed rumor detection models highly depend on manually designed features. These features are mainly concentrated in	text content and users' information.

For example, Castillo et al.~\cite{castillo2011information} design various types of features (e.g., sentence length, number of sentiment words, user's age, number of followers and friends) to evaluate the credibility of a message on specific topics. Yang et al.~\cite{yang2012automatic} introduced additional client-based and location-based features to identify rumors in Sina Weibo. Kwon et al.~\cite{kwon2013prominent} employed temporal, structural and linguistic features to improve the performance of rumor detection. Liu et al.~\cite{liu2015real} proposed verification features by treating the crowd's contradictory beliefs as their debates on veracity. Ma et al.~\cite{ma2015detect} integrated the temporal characteristics of topical features and sentiment features into rumor detection. Wu et al.~\cite{wu2015false} proposed to capture the high-order propagation patterns to improve rumor detection. 

Most of these feature-based methods are biased, time-consuming and limited. They are usually designed for specific scenarios and hence cannot be easily generalized for other applications.

\subsection{Deep Neural Networks Related Methods}
To address above issues of traditional feature-based methods, researchers employed deep neural networks to learn efficient features automatically for rumor detection. 

Ma et al.~\cite{ma2016detecting} utilized various recurrent neural networks (RNN) to model the repost sequence. Yu et al.~\cite{yu2017convolutional} employed convolutional neural networks (CNN) on the repost sequence to capture the interactions among high-level features. 

There is also a technical trend mentioned earlier that combines neural network with manual design features roundly. Bhatt et al.~\cite{bhatt2018combining} combined the neural, statistical and external features using deep MLP.  Ruchansky et al. ~\cite{ruchansky2017csi} proposed a model that combines three characteristics: the text of an article, the users' response, and the source users' information promoting. 

But this method of combining different information is only an increase in the types of information that can be used, not paying enough attention to early detection. Moreover, all these neural network related models can only detect rumors with the consideration of the entire or fixed proportions of repost information, which is not able to detect rumors as early as possible in practice. 

\subsection{Propagation Mode Related Method}
Unlike the previous two kinds of methods which focus on the use of the text content information, the propagation mode related methods focus on the differences in the characteristics of real and false information transmission. 

Vosoughi et al.~\cite{vosoughi2018spread} verified that there is a substantial difference in the actual transmission of false information and real information by studying a large number of Twitter data experiments. Falsehood diffused significantly farther, faster, more in-depth, and more broadly than the truth in all categories of information. Hoaxy et al.~\cite{shao2016hoaxy} found that the sharing of fact-checking content typically lags that of misinformation by 10 $\sim$ 20 hours. Moreover, fake news is dominated by very active users, while fact checking is a more grassroots activity. Ma et al.~\cite{ma2017detect} proposed a kernel-based method, which captures high-order patterns differentiating different types of rumors by evaluating the similarities between their propagation tree structures. 

However, the rumor detection methods based on propagation model have not yet been fully developed. So far no studies have shown whether the difference in the mode of transmission in the early dissemination of rumors helps the detection of rumors. It is necessary to explore the characteristics of essential propagation modes to achieve more reliable rumor early detection.

\vspace{3mm}

There have been some early rumor detection methods in recent years~\cite{zhao2015enquiring,nguyen2017early,wu2017gleaning}. Nevertheless, similar to feature-based methods, most existing models heavily depend on well-designed features or signals in repost or comment information.

Thus in this paper, inspired by the mode of propagation, we take advantage of deep neural networks on feature learning and propose our early rumor detection model CED. The effectiveness of CED has been verified through experiments on real-world datasets.

\section{Methodology}
In this section, we will introduce how to classify a microblog into rumors or facts according to its repost sequence. First, we will formalize the early rumor detection problem and give essential definitions. Then we will introduce basic concepts of recurrent neural networks. Finally, we will describe the details of our proposed model for early rumor detection.

\subsection{Problem Formalization}
Microblogs in social media platforms are usually conditioned to a certain length and thus contain limited information for rumor detection. Therefore, we utilize the relevant repost information for identification as the previous work~\cite{ma2016detecting} did.

Assume we have a set of microblogs as $\mathbf{M}=\{M\}$. Each original microblog message $M$ has a relevant repost sequence, denoted as $M=\{(m_i, t_i)\}$. Here, each repost $m_i$ represents the text content, and $t_{i}$ denotes the corresponding time stamp. Generally, rumor detection aims to predict the label $y\in \{0, 1\}$ for each microblog $M$, according to the repost sequence $\{(m_i, t_i)\}$. Note that, we set $y=1$ for rumors and $y=0$ otherwise.

\subsection{Repost Sequence Partition}
Following the practice in~\cite{ma2016detecting}, we need to convert original repost sequences into more informative neural network inputs because of two major reasons. On the one hand, most reposts reserve no critical information and will not benefit rumor prediction. It is very common that people repost a microblog without expressing any comments or opinions. On the other hand, a large number of microblogs own thousands of reposts, which can not be well addressed by existing neural networks and will also introduce computational efficiency issues.

Therefore, we convert the original repost sequence of a microblog by merging constant  reposts into a single unit. Specifically, we batch a certain number ($N$) of consecutive reposts together. To ensure each batch of reposts is informative and the time granularity is small enough, we set $N=10$ practically. After this processing, the repost sequence of $M=\{(m_i, t_i)\}$ is converted to $F=\{f_i\}$, where $f_i=\{m_{(i-1)*N+1}, \cdots ,m_{i*N}\}$.

\subsection{Feature Acquisition of Parted Repost Sequence}
\label{CNNs}
Based on the converted sequence, we need to transform the text information in each interval into a feature vector and feed this vector into neural networks. We have two ways to get feature vectors $\mathbf{F}$ for every converted repost sequence $F$ as follows: 

\subsubsection{TF-IDF}
Following~\cite{ma2016detecting}, we employ the simple and efficient text representation method TF-IDF~\cite{salton1988term} for interval representation. Assuming the vocabulary size is $V$, we can obtain a variable-length matrix $\mathbf{F} = \{x_i\} \in  \mathbb{R}^{V\times |F|}$ for each converted repost sequence $F$. 

\subsubsection{Convolutional Neural Network}
Convolutional Neural Network(CNN) has been successfully applied in sentence semantic analysis ~\cite{kalchbrenner2014convolutional}, click-through rate prediction ~\cite{liu2015convolutional}, text classification ~\cite{kim2014convolutional} and reinforcement learning tasks ~\cite{tamar2016value}, which is made up of stacked convolutional and pooling layers, the architectures of which help model significant semantic features and achieve much improvement in respective fields. CNN is usually trained through stochastic gradient descent (SGD), with back-propagation to compute gradients.

To get the feature vector of each departed repost $f_i =\{w_1, \ldots, w_k\}$ of converted repost sequence $F=\{f_i\}$, we first represent each word $w_i$ with its word embeddings vector $\mathbf{w}_i\in \mathbb{R}^{d}$, where $d$ is the dimension of word embeddings. Then each departed repost $f_i$ of length n (padded when necessary) is represented as instance matrix:
\begin{equation}
\small
\label{eq:cnn1}
\mathbf{w}_{1:n} = \mathbf{w}_{1} \oplus \mathbf{w}_{2} \oplus \cdots \oplus \mathbf{w}_{n}
\end{equation}
where $\oplus$ is the concatenation operator, so $\mathbf{w}_{1:n} \in \mathbb{R}^{d \times n}$. A convolutional layer is obtained by convolution operations of a weight matrix $G \in \mathbb{R}^{d \times h}$ in a  row-wise way, where $h$ is the length of words window applied to produce a new feature. Followed by a nonlinearity function $ReLU$ ~\cite{glorot2011deep} applied to the convolution result, an element of a feature map can be obtained as: 
\begin{equation}
\small
\label{eq:cnn2}
c_i = ReLU( {< G, \mathbf{w}_{i:i+h-1}>}_{\mathcal{F}} + b) 
\end{equation}
note that $c_i \in \mathbb{R}^{n-h+1}$, where $\mathbf{w}_{i:i+h-1}$ is the $i$-th to $(i+h-1)$-th columns of $\mathbf{w}_{1:n}$, $b$ is bias term and the subscript $\mathcal{F}$ is the Frobenius inner product, i.e., the summation of products of corresponding elements of both matrices. We then apply a max-pooling operation over the feature map $c$ and take the maximum value $\hat{c} = max{(c)}$ as the most significant feature corresponding to the weight matrix $G$. Now we have described the process how the feature is extracted from one filter now. 

The CNN model uses multiple filters (with varying window sizes and different values in weight matrix) to obtain multiple features. Assume that the length of the feature vector obtained from CNN for each departed repost $f_i$ is E, we can get a variable-length matrix $\mathbf{F} = \{x_i\} \in  \mathbb{R}^{E\times |F|}$ for each converted repost sequence $F$. Moreover, sometimes the above-mentioned layer could be repeated to yield deeper layers to attain high-level interactions features. These features can be passed to a fully connected softmax layer whose output is the probability distribution over labels, but in our model feature matrix, $\mathbf{F}$ of repost sequence will be sent into RNN as input.

\subsection{Rumor Detection with RNNs}
Recurrent neural networks (RNNs) are a typical class of feed-forward neural networks for sequence modeling and have achieved great success in natural language processing tasks, such as text classification and machine translation. Generally, RNNs deal with variable-length sequences through a recurrent unit. For the $i$-th step, recurrent unit updates its hidden state $h_i$ based on previous hidden state $h_{i-1}$ and current input $x_i$.

Specifically, Cho et al.~\cite{cho2014properties} propose Gated Recurrent Unit (GRU), which involves two gates, i.e., the reset gate $r$ and update gate $z$.  The calculation of the hidden state in GRU is as follows:
	\begin{equation}
	\small
	\begin{split}
	r_i &= \sigma (U_R \cdot x_i +W_R\cdot  h_{i-1} ), \\
	z_i &= \sigma (U_Z \cdot x_i+  W_Z \cdot h_{i-1}), \\
	\widetilde{h}_i &= \tanh (U_H\cdot x_i +  W_H\cdot (h_{i-1} \odot r_i)), \\
	h_i &= (1 - z_i) \cdot h_{i-1} + z_{i} \cdot \widetilde{h}_i. \\
	\end{split}
	\end{equation}
	Here, $\odot$ indicates the element-wise multiplication. $U$ and $W$ are the weight matrices.

After feeding the feature matrix $\mathbf{F}$ of a repost sequence into various RNNs, we can get every hidden state of each interval and take the final hidden state $h_{|F|}$ as the representative vector of repost sequence.

The objective function of rumor detection is the log-likelihood of predicting the label $y \in \{0,1\}$ given the representation vector as follows:
\begin{equation}
\small
\label{eq:lld}
\mathcal{O}(|F|) = \log p(y|h_{|F|}),
\end{equation}
where $p(1|h_i) = \sigma (h_i \cdot s)$ and $p(0|h_i) = 1- p(1|h_i)$. Here, $s$ is weight vector and will be learnt during training.

\subsection{Early Rumor Detection}~\label{ER}
Early rumor detection aims to make a credible prediction for a microblog at an advanced time point, while existing rumor detection methods can only predict considering the entire repost information.

\subsubsection{\textbf{Credible Early Detection (CED)}}
\begin{figure}[!htb]
	\centering
  	\includegraphics[width=0.9\columnwidth]{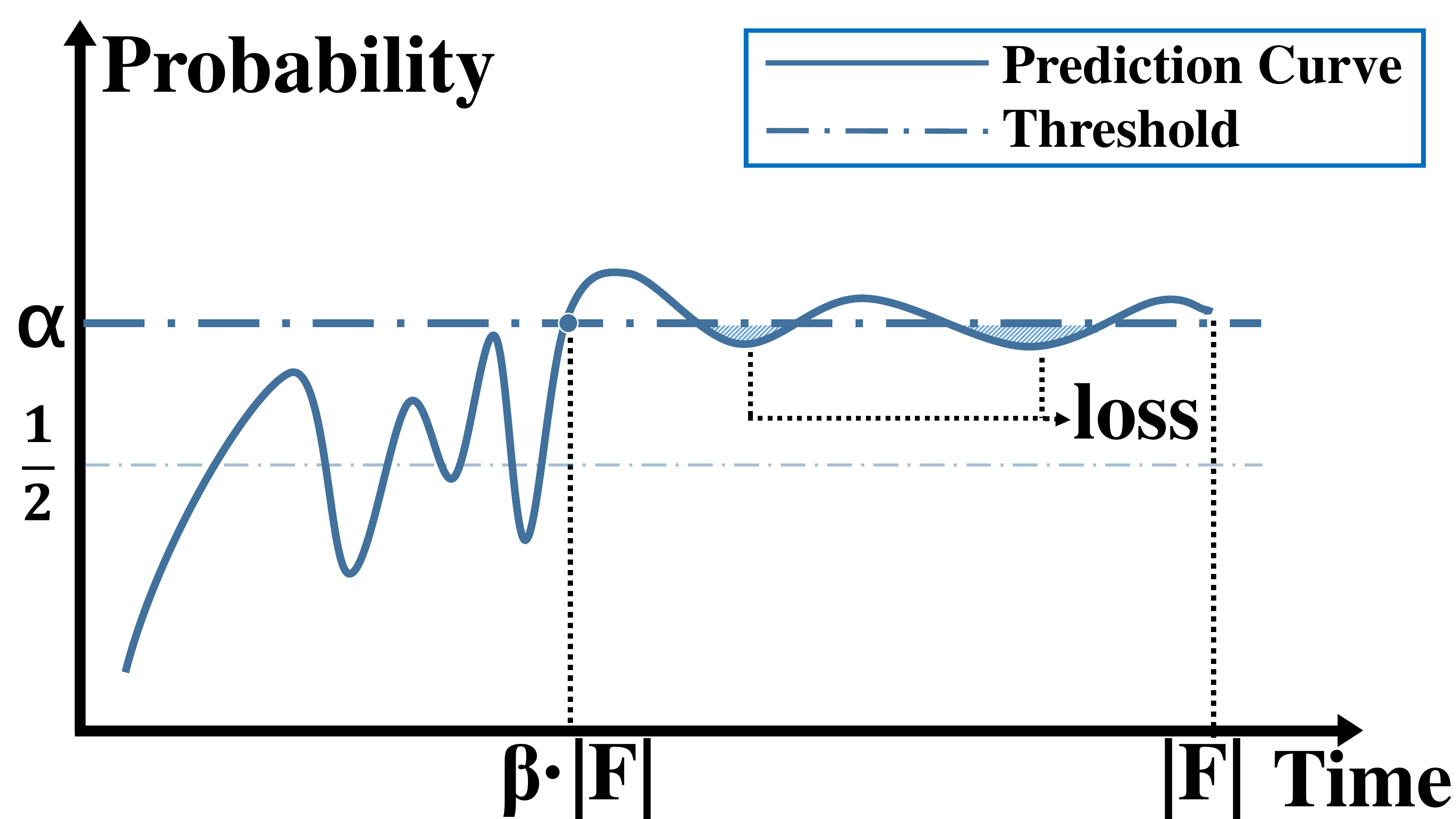}
	\caption{An illustration of the prediction curve. Credible Detection Point shows at $\beta|\mathbf{F}|$ time point.}
	\label{fig:diff}
\end{figure}
According to our observation over the repost information, we assume that there exists a ``Credible Detection Point'' for each microblog. As illustrated in Fig.~\ref{fig:diff}, before this time point, there usually exist conflict reposts that support or oppose the original microblog, which will confuse people. During this period, the rumors are difficult to distinguish for both people and rumor detection models. The predictions of rumor detection models are unstable which may disturb the performance of early rumor detection models. However, after this time stamp, the debate about the original microblog reaches an agreement, and the following prediction should be stable and credible.

Base on this assumption, we introduce a parameter $\beta \in [0, 1]$ for each microblog to represent the ``Credible Detection Point''. This parameter $\beta$ is determined by the time point when the prediction threshold is reached for the first time:
\begin{equation}
\small
\label{eq:beta}
\beta = \frac{n_f}{|F|}
\end{equation}
where $n_f$ is the time point when prediction probability reaches the threshold $\alpha$, i.e.,  $p(1|h_{n_f}) \ge \alpha$ or $p(0|h_{n_f}) \ge \alpha$ for the first time.

Thus, the first part of our objective aims to maximize the prediction probability after this time point as follows:
\begin{equation}
\small
\label{eq:predict}
\mathcal{O}_{pred} = \frac{1}{(1-\beta)|F|}\sum_{\beta|F| \le i \le |F|} \mathcal{O}(i).
\end{equation}

To ensure the property of early detection, the second part of our objective aims to minimize the prediction time $\beta$  as follows:
\begin{equation}
\small
\label{eq:time}
\mathcal{O}_{time} = -\log \beta.
\end{equation}

Besides, we also want to guarantee the reliability of the detection point. In other words, the prediction probabilities after this detection point should be stable and scale out the threshold. As shown in Fig.~\ref{fig:diff}, for $y=1$, we aim to minimize the difference value under the upper threshold, and for $y=0$, we want to minimize the difference exceed the lower threshold. The objective function is as follows:
\begin{equation}
\small
\label{eq:diff}
\begin{split}
\mathcal{O}_{diff} &=  -\frac{1}{(1-\beta) |F|}\sum_{\beta |F| \le i \le |F|} (y\cdot \max(0, \log \alpha \\ - \mathcal{O}(i))  
&+ (1-y) \cdot \max(0, \mathcal{O}(i) - \log(1 - \alpha))).
\end{split}
\end{equation}
At last, we introduce two hyper-parameters $\lambda_0$ and $\lambda_1$ to combine these three objectives as follows:
\begin{equation}
\small
\label{eq:model2}
\mathcal{O}_{CED} = \mathcal{O}_{pred}+\lambda_0\cdot \mathcal{O}_{diff}+\lambda_1\cdot \mathcal{O}_{time}.
\end{equation}

\begin{figure*}[!t]
\centering
\includegraphics[width=0.8\textwidth]{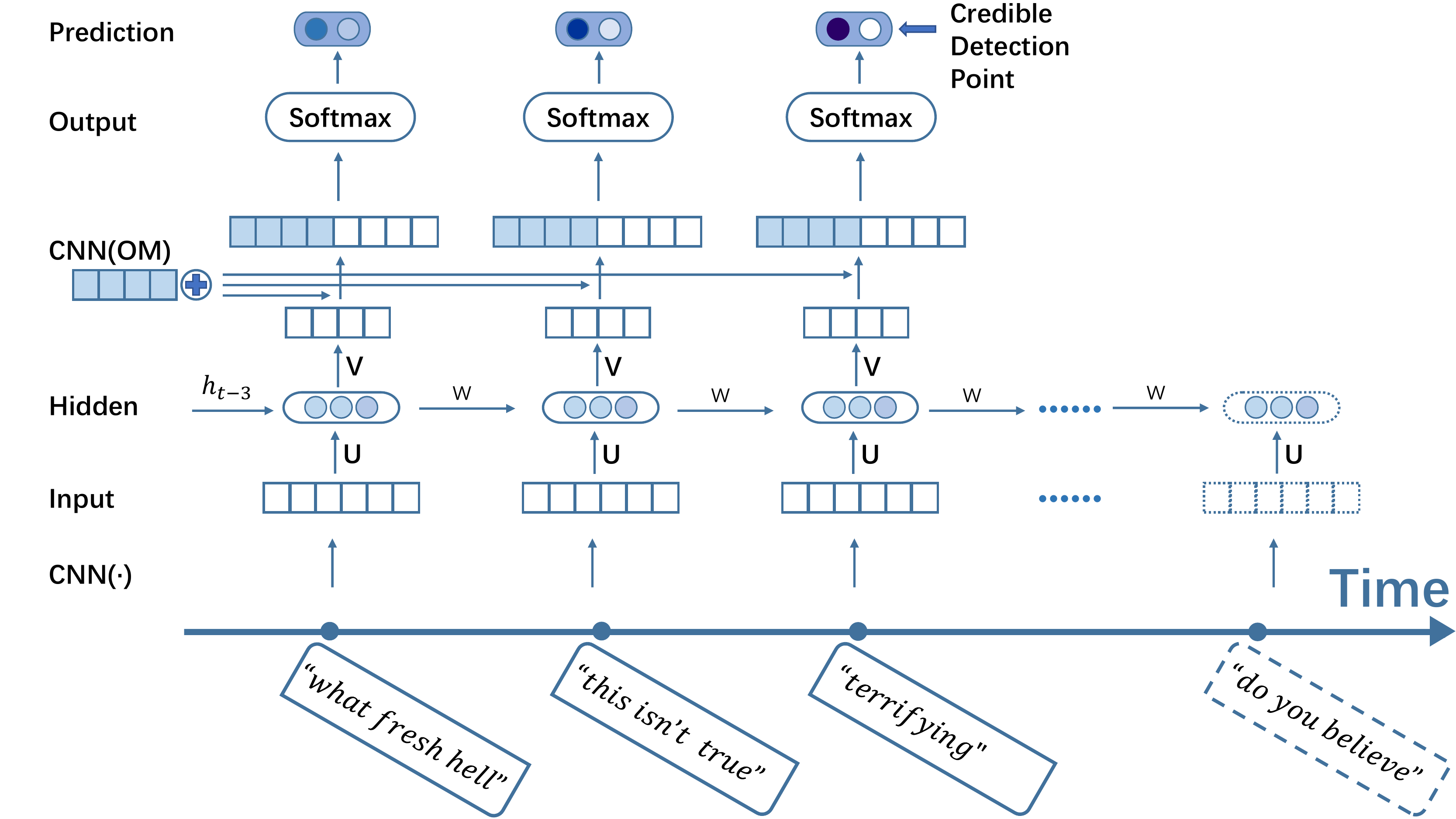}
\caption{Model structure of CED-CNN. $CNN(\cdot)$ refers to the CNN method to obtain feature vector as mentioned in section~\ref{CNNs}. Based on this, $CNN(OM)$ means the feature vector got from original microblog. As the repost information is continuously sent to the model in chronological order, the prediction probability gap is getting larger. Once Credible Detection Point is obtained, the subsequent information is no longer needed, which is indicated by dotted lines.}
\label{fig:ced_cnn}
\end{figure*}

\subsubsection{\textbf{CED with Original Microblog}}

Original microblog messages play an important role in rumor detection, which is usually regarded as a specific repost by existing repost-based rumor detection models. In this part, we extend CED to integrate original microblog information in a simple way, denoted as \textbf{CED-OM}.

In CED-OM, we apply Convolutional Neural Network(CNN) with commonly used architecture as in~\cite{kim2014convolutional} to get the feature vector of original microblogs. Specifically, for each original microblog $\mathbf{m}=\{w_1, \ldots, w_k\}$, we first convert each word $w_i$ into its word embeddings $\mathbf{w}_i\in \mathbb{R}^{d}$, where $d$ is the dimension of word embeddings. Afterwards, we apply convolution operation to the $i$-th sliding window of length $h$ through $\mathbf{c}_i = G\cdot \mathbf{w}_{i:i+h-1} + b$, where $G$ and $b$ are convolution matrix and bias vector respectively. At last, we apply a max-pooling operation over $\mathbf{c}_i$ to obtain the feature $r$. Then we get the final feature vector $\mathbf{r}$ by controlling the number of convolution operations with varying window sizes $h$ and different values in weight matrix.

We simply concatenate the feature vector $\mathbf{r}$ with each hidden state $h_i$ to get the final hidden state as $\hat{h}_i=h_i\oplus \mathbf{r}$ getting larger vector dimension, which is further used to replace  $ \mathcal{O}(i) = \log p(y|h_i)$ in Eq. \ref{eq:model2}. Note that, we employ a simple way to incorporate original microblog into rumor detection. More complex and effective approaches, such as attention mechanism~\cite{dos2016attentive,tu2017cane}, can be introduced to enhance the microblog representations, which will be explored in future work.

To optimize the objective in Eq. \ref{eq:model2}, we employ an efficient optimization algorithm Adam~\cite{kingma2014adam}. Note that, the microblog-specific parameter $\beta$ will also be determined by Eq. \ref{eq:beta} during training.

\subsubsection{\textbf{CED using CNN to deal with reposts}}
On the basis of the previous method using original microblog, we proposed a more effective method to handle the reposts sequence as mentioned in section~\ref{CNNs}, that is, using CNN instead of TF-IDT. In this part, we propose \textbf{CED-CNN} to obtain feature vectors of repost sequence more efficiently, achieving much more excellence \textbf{Early detection}.

Analogous to CED, we segment the reposts sequence. At the same time, we set the maximum length $L$ of the repost sequence $M=\{(m_i, t_i)\}$ by the observing statistical result. For each microblog, repost sequence gets the same length $L$ through padding, while real length $|F| \leq L$. Each repost $m_i = \{w_1, \ldots, w_k\}$ of every microblog consists indefinite-length words. We need to convert each word $w_i$ into word vector and pad each repost into same sentence length. Then we get word vector representation $\mathbf{w}_{1:n}$ of each repost, where $n$ is the padding length of the repost sentence.
\begin{equation}
\small
\label{eq:ced_cnn1}
\mathbf{w}_{1:n} = \mathbf{w}_{1} \oplus \mathbf{w}_{2} \oplus \cdots \oplus \mathbf{w}_{n}
\end{equation}
We note the CNN method to obtain feature vectors of reposts sequence, which is mentioned in section~\ref{CNNs}, as $CNN(\cdot)$, the final output feature map is obtained as fellow:
\begin{equation}
\small
\label{eq:ced_cnn2}
\mathbf{F} = \{x_i\} = CNN( \mathbf{w}_{1:n} )
\end{equation}
where the length of the feature vector obtained from CNN for each departed repost $f_i$ is E, and $\mathbf{F} \in  \mathbb{R}^{E\times |F|}$ for each converted repost sequence $F$. 

Similar to CED and CED-OM, we send this feature sequence $\mathbf{F}$ to RNN. At the same time, we apply CNN with commonly used architecture to get the feature vector of original microblog, noted as $\mathbf{r}$. Then We concatenate the feature vector $\mathbf{r}$ with each hidden state $h_i$ to get the final hidden state as $\hat{h}_i=h_i \oplus \mathbf{r}$, which is further used to replace  $ \mathcal{O}(i) = \log p(y|h_i)$ in Eq. \ref{eq:model2}. 

To sum up, we hire CNN to obtain feature vector of original microblog $\mathbf{r}$ and feature vectors of repost sequence $\mathbf{F}$ in the CED-CNN method, which is shown in Fig.~\ref{fig:ced_cnn}. Departed repost feature map $\mathbf{F}$ is set to RNN as input, one column each time. After getting each hidden state, we use the original microblog feature vector $\mathbf{r}$ to expand the dimension of each hidden state and get prediction probability, where the objective function is same as that in Eq. \ref{eq:model2}. The early detection point is determined also as shown in Eq. \ref{eq:beta}. During the training process, we check the prediction probabilities at every step of report sequence. When it reaches the threshold for the first time, we get parameter $\beta$ of current repost sequence and calculate the objective function based on $\beta$. It is worth pointing out that the parameter $\beta$ represents the ``Credible Detection Point'' during testing.

\subsubsection{Detection Strategy for Testing}
To realize the early detection of rumors during testing, we propose a threshold based rumor detection strategy. Sequentially, we calculate the prediction probabilities $p_i = p(1|h_i)$ at each step of the repost sequence and make a prediction as follows:
\begin{equation}
\small
\widetilde{y}=
\begin{cases}
1, &\mbox{if $p_i \ge \alpha$},\\
0, &\mbox{if $p_i < 1-\alpha$},\\
\phi, &\mbox{otherwise}.
\end{cases}
\end{equation}
Here, $\alpha\in [0.5, 1]$ is a pre-defined threshold as in Eq. \ref{eq:diff}. if $\widetilde{y} \in \{0,1\}$, it means the prediction probability reaches the threshold and we can make a credible prediction only using $i$ intervals rather than $|F|$. If $\widetilde{y} = \phi$, we will look for the prediction result of next step.

\section{Experiments}
To verify the effectiveness of our proposed early detection models, we perform rumor detection experiments on three representative real-world datasets including Weibo and Twitter. Besides, we also conduct detailed analysis on detection accuracy, early detection, and parameter sensitivity.

\begin{table}[!htb]
	\caption{Statistics of the datasets. (Here, \emph{Rep.} indicates reposts.)}
	\label{table:datasets}
	\small
	\begin{center}
		\makebox[0.6\linewidth][c]{
		\begin{tabular}{c|r|r|r}
			\toprule
			Weibo-all & Rumors & Non-rumors & All \\ \midrule
			Numbers & $3,851$ & $4,199$ & $8,050$ \\
			Reposts & $2,572,047$ & $2,450,821$ & $5,022,868$ \\
			Min. Rep. & $3$ & $5$ & $3$ \\
			Max Rep.& $59,317$ & $52,156$ & $59,317$ \\
			Ave. Rep. & $668$ & $584$ & $624$ \\ \midrule\midrule
			Weibo-stan & Rumors & Non-rumors & All \\ \midrule
			Numbers & $2,313$ & $2,350$ & $4,663$ \\
			Reposts & $2,088,430$ & $1,659,258$ & $3,747,688$ \\
			Min. Rep. & $3$ & $5$ & $3$ \\
			Max Rep.& $59,317$ & $52,156$ & $59,317$ \\
			Ave. Rep. & $903$ & $706$ & $804$ \\ \midrule\midrule
			Twitter & Rumors & Non-rumors & All \\ \midrule
			Numbers & $495$ & $493$ & $988$ \\
			Reposts & $148,594$ & $397,373$ & $545,967$ \\
			Min. Rep. & $4$ & $5$ & $4$ \\
			Max Rep.& $7,071$ & $37,087$ & $37,087$ \\
			Ave. Rep. & $300$ & $806$ & $553$ \\
			\bottomrule
	\end{tabular}}
    \end{center}
\end{table}

\subsection{Datasets}
For evaluation, we employ two standard real-world rumor datasets from Sina Weibo\footnote{https://www.weibo.com/} and Twitter\footnote{https://twitter.com} as in~\cite{yu2017convolutional}, which is separately labeled as ``Weibo-stan'' and ``Twitter''. We get the microblog ID lists of the two datasets and collect all the repost information of each microblog. Note that, some microblogs are unavailable as they are deleted or their belonging accounts are closed. 

Also, in order to obtain a larger dataset to verify the effectiveness of our model, we also model on Ma et al.~\cite{ma2016detecting} method to crawl more Weibo data to get a larger dataset, which is noted as ``Weibo-all''. For more Weibo data, we obtain a set of known rumors from the Sina community management center\footnote{http://service.account.weibo.com}, which reports various misinformation. The Weibo API can capture the original messages and all their repost/reply messages given. We also gather a similar number of non-rumor events by crawling the posts of general threads that are not reported as rumors on Sina community management center. The detailed statistics of these three datasets are listed in Table~\ref{table:datasets}.

\subsection{Baselines}
To detect rumors according to the repost information, we employ four representative baselines as follows:

\textbf{CNN-OM~\cite{kim2014convolutional}} CNN based models have achieved promising results in text classification tasks. Here, we employ CNN in~\cite{kim2014convolutional} to deal with just the original microblogs for rumor classification.

\textbf{TF-IDF~\cite{salton1988term}} We simply merge all the text information of the repost sequence into the bag of words and calculate the TF-IDF representation vector of this sequence. With these TF-IDF representations, we train an SVM~\cite{suykens1999least} classifier for rumor classification.

\textbf{GRU-2~\cite{ma2016detecting}} Ma et al. propose to utilize RNNs to learn representations of repost sequences. Here, we follow this idea and employ a $2$-layer GRU~\cite{cho2014properties} to train rumor classifiers.
	
\textbf{CAMI~\cite{yu2017convolutional}} Yu et al. 2017 employ paragraph vector to represent each repost internal and utilize CNN to capture high-level interactions for misinformation identification\footnote{As the source code of this work is not available, we reproduce this model with the help of the authors.}.

\begin{table*}[!htb]
	\caption{Weibo experimental results. (0.875 and 0.975 are prediction threshold of CED, CED-OM and CED-CNN.)}
	\label{table:result-1}
	\small
	\begin{center}
	\makebox[0.7\linewidth][c]{
	\begin{tabular}{c|c|c|c|c|c|c|c|c|c|c}
		\toprule
		& \multicolumn{5}{|c|}{Weibo-stan} & \multicolumn{5}{|c}{Weibo-all} \\ \midrule
		Methods & Acc. & Precision & Recall & F1 & ER & Acc. & Precision & Recall & F1 & ER\\ \midrule\midrule
		CNN-OM & $0.809$ & $0.829$ & $0.781$ & $0.804$  & $100\%$ & $0.887$ & $0.883$ & $0.884$ & $0.883$ & $100\%$\\ \midrule
		TF-IDF & $0.859$ & $0.799$ & $0.951$ & $0.868$ & $100\%$ & $0.819$ & $0.914$ & $0.678$ & $0.779$ & $100\%$ \\ \midrule
		GRU-2 & $0.920$ & $0.926$ & $0.900$ & $0.913$ & $100\%$ & $0.906$ & $0.923$ & $0.878$ & $0.901$ & $100\%$\\ \midrule
		CAMI & $0.896$ & $0.890$ & $0.876$ & $0.883$ & $100\%$ & $0.893$ &$0.866$ & $0.914$ & $0.889$ & $100\%$\\ \midrule
		CED(0.875) & $0.938$ & $0.930$ & $0.946$ & $0.938$ & $32.7\%$ & $0.913$ & $0.926$ & $0.892$ & $0.908$ & $31.3\%$ \\ 
		CED(0.975) & $\mathbf{0.946}$ & $\mathbf{0.946}$ & $0.944$ & $\mathbf{0.945}$  & $41.4\%$ & $0.921$ & $0.934$ & $0.899$ & $0.916$ & $45.6\%$\\ \midrule
		CED-OM(0.875) & $0.916$ & $0.896$ & $0.941$ & $0.918$ & $28.1\%$ & $0.920$ & $0.930$ & $0.902$ & $0.916$ & $19.6\%$ \\
		CED-OM(0.975) & $0.942$ & $0.923$ & $\mathbf{0.964}$ & $\mathbf{0.943}$ & $32.0\%$  & $0.913$ & $0.942$ & $0.874$ & $0.906$ & $19.5\%$ \\ \midrule
        CED-CNN(0.875) & $0.900$ & $0.920$ & $0.889$ & $0.904$ & $\mathbf{15.1}\%$ & $0.941$ & $0.947$ & $0.929$ & $0.938$ & $\textbf{13.2}\%$ \\ 
		CED-CNN(0.975) & $0.912$ & $0.920$ & $0.912$ & $0.916$  & $19.3\%$ & $\textbf{0.947}$ & $\textbf{0.954}$ & $\textbf{0.934}$ & $\textbf{0.944}$ & $17.9\%$\\ \midrule
		\bottomrule
	\end{tabular}}
	\end{center}
\end{table*}

\begin{table}[!t]
\renewcommand{\arraystretch}{1.3}
\caption{Twitter Experimental results.}
\label{table:result-2}
\centering
\begin{tabular}{c|r|r|r|r|r}
			\toprule
			Methods & Acc. & Precision & Recall & F1 & ER\\ \midrule\midrule
			TF-IDF & $0.587$ & $0.569$ & $0.941$ & $0.710$ & $100\%$ \\ \midrule
			GRU-2 & $0.672$ & $0.626$ & $0.779$ & $0.694$ & $100\%$ \\ \midrule
			CAMI & $0.595$ & $0.667$ & $0.525$ & $0.588$ & $100\%$ \\ \midrule
			CED(0.875) & $0.717$ & $0.692$ & $0.733$ & $0.712$ & $53.0\%$ \\ 
			CED(0.975) & $\mathbf{0.744}$ & $\mathbf{0.708}$ & $0.791$ & $0.747$ & $52.5\%$ \\ \midrule
            CED-CNN(0.875) & $0.721$ & $0.642$ & $0.929$ & $0.760$ & $\mathbf{32.1\%}$ \\
			CED-CNN(0.975) & $0.704$ & $0.616$ & $\mathbf{1.000}$ & $\mathbf{0.762}$ & $43.8\%$ \\
			\bottomrule
	\end{tabular}
\end{table}

\subsection{Evaluation Metrics and Parameter Settings}

To evaluate the performance of various methods on rumor detection, we adopt \emph{accuracy} (Acc.) metric as follows:

\begin{equation}
\small
accuracy = \frac{a}{T},
\end{equation}
where $a$ is the number of correctly predicted rumors and non-rumors in testing set , and $T$ is the size of the testing set. Meanwhile, we also evaluate the performance of these models using macro \emph{precision}, \emph{recall} and \emph{F-measure} metrics.

Besides, we also want to demonstrate the effectiveness of our proposed early detection model. Thus, we propose \emph{Early Rate} (ER) to evaluate the utilization ratio of repost information as follows:
\begin{equation}
\small
ER = \frac{1}{T}\sum_{i \in Testing}\frac{t_i}{|F_i|},
\end{equation}
where $t_i$ indicates that the model makes a prediction at $t_i$ time stamp for the $i$-th microblog, i.e, prediction probability reaches the threshold $\alpha$ for the first time, and $|F_i|$ is the total length of intervals of the $i$-th repost sequence. Lower \emph{ER} value means the model can detect rumors earlier, with less repost information used.

Following the settings in~\cite{ma2016detecting}, we randomly select $10\%$ instances as the validation dataset, and split the rest for training and testing with a ratio of $3:1$ in all three datasets (Weibo-stan, Weibo-all and Twitter). We set the dimension of  TF-IDF vector to $1000$, the vocabulary size of original microblogs to $3,000$ when using CNN-OM and CED-CNN and the collection vocabulary size of original microblogs and repost sequences to $20,000$.

We show the results of CED under two settings of prediction threshold $\alpha$, including $0.875$ and $0.975$. The weights $\lambda_0$ and $\lambda_1$ are set to $0.01$ and $0.2$ respectively. Besides, we employ the same neural network settings as~\cite{ma2016detecting} and utilize a $2$-layer GRU as sequence encoder.

For a fair comparison, we set the hidden size to $200$ for CED, GRU-2 and CAMI, and $100$ for CED-OM and CED-CNN. For CNN part of original microblog in CED-OM and CED-CNN, we set the filter widths to $[4, 5]$, with each filter size to $50$. As a result, the final hidden state in CED-OM and CED-CNN is $200$ as well.

\subsection{Results and Analysis}
As shown in Table~\ref{table:result-1} and Table~\ref{table:result-2}, we list the detailed results of different methods under various evaluation metrics. We also bold the best result of each column in both tables. Note that, the original microblogs in the Twitter dataset are not released by~\cite{ma2016detecting,yu2017convolutional}. Therefore, we omit the results of CNN-OM and CED-OM on this dataset. From the experimental results, we have the following observations:

(1) Our proposed early rumor detection models: CED, CED-OM, and CED-CNN, achieve significant and consistent improvements compared with all baseline methods, with higher accuracy and even much less repost sequence information. It demonstrates the reasonability and effectiveness of our proposed early detection models and strategy. 

(2) Comparing with all the baselines, our model can considerably reduce the detection time by around $86\%$ in Weibo and $77\%$ in Twitter, with even better performance on accuracy. The reason is that CED aims to make the representation of each single repost intervals predictable. On the contrary, other methods only make the final representation predictable. Such practice makes our model capable of detecting early with partial repost information. Especially, CED-CNN achieves a very high detection accuracy using only $13.2\%$ of the repost sequence information, making it possible to detect rumors in real-time on social media.
	
(3) By incorporating original microblogs, CED-OM achieves significant improvement on early rate under both thresholds. It is mainly because that original microblogs can provide additional information when there are only a few of reposts. The lack of reposts raises difficulties for CED to detect rumors, which can be addressed by considering original microblogs to some extent.

(4) The Early Rate of CED-CNN method achieves the best performance in all three data sets. At the same time, it also has excellent performance in Accuracy, especially after being fully trained in a large dataset like Weibo-all. Compared with CED, CED-CNN has better performance in Early Rate performance, which shows that the method using CNN to extract the characteristics of forwarding sequence information is more efficient than TF-IDF.

(5) The detection threshold $\alpha$ plays an important role in balancing the detection accuracy and early rate. As shown in Table~\ref{table:result-1}, a higher threshold means that our model will give more confident prediction result, while delaying the detection time and resulting in a higher early rate. The threshold provides a flexible selection according to the actual scenarios. 
	
Observations above demonstrate that our model is able to make a more accurate prediction with limited repost information. It is robust and flexible to various datasets and parameter settings.

\begin{table*}[!htb]
\renewcommand{\arraystretch}{1.3}
\caption{Loss experimental results on Weibo-stan dataset.}
\label{table:losscompare}
\centering
\begin{tabular}{c|r|r|r|r|r}
			\toprule
			Methods & Acc. & Precision & Recall & F1 & ER\\ \midrule\midrule
            L1-CED-CNN(0.975) & $0.949$ & $0.961$ & $0.930$ & $0.946$ & $21.1\%$ \\ 
            L2-CED-CNN(0.975) & $0.934$ & $0.916$ & $0.948$ & $0.932$ & $20.8\%$ \\ 
			CED-CNN(0.975) & $0.947$ & $0.954$ & $0.934$ & $0.944$ & $17.9\%$\\
			\bottomrule
	\end{tabular}
\end{table*}

\subsection{Early Detection.}
In CED, the ``Credible Detection Point'' for each microblog is determined by Eq. \ref{eq:beta} and constantly ahead during training and inferred during testing based on threshold-based detection strategy. In other words, CED can learn and infer an appropriate detection point for various microblogs.

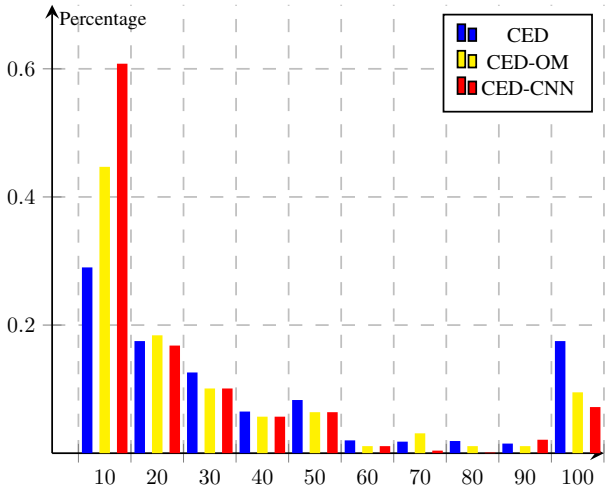
\begin{figure}[!htb]
\centering
        \resizebox{0.45\textwidth}{!}{
			\begin{tikzpicture}
			\begin{axis}[
			ylabel = {\small Percentage},
			ymax=0.7,
			enlargelimits=0.01,
			ybar interval=0.6,
			axis lines=center,
			ymajorgrids=true,
			grid style=dashed,
            legend image post style={scale=0.5},
            nodes={scale=0.5, transform shape},
			]
			\addplot[draw=none,fill=blue] coordinates {
				(5, 0)(10, 0.290)(20, 0.175)(30, 0.126)(40, 0.065)(50, 0.083)(60, 0.020)(70, 0.018)(80, 0.019)(90, 0.015)(100, 0.175)(110, 0)
			};
            
            \addplot[draw=none,fill=yellow] coordinates {
				(5, 0)(10, 0.447)(20, 0.184)(30, 0.101)(40, 0.057)(50, 0.064)(60, 0.011)(70, 0.031)(80, 0.011)(90, 0.011)(100, 0.095)(110, 0)
			};  
			
			\addplot[draw=none,fill=red] coordinates {
				(5, 0)(10, 0.608)(20, 0.168)(30, 0.101)(40, 0.057)(50, 0.064)(60, 0.011)(70, 0.004)(80, 0.001)(90, 0.021)(100, 0.072)(110, 0)
			};    
			
			\legend{CED, CED-OM,CED-CNN}
			\end{axis}
			\end{tikzpicture}}
	\caption{Early rate distribution in Weibo-all dataset. The abscissa represents the percentage of repost sequence information used to reach Credible Detection Point(\%). }
	\label{fig:distribution}
\end{figure}

In Fig.~\ref{fig:distribution}, we show the distribution of ``Credible Detection Point'' in testing set. In order to reflect the specific performance of different models in different datasets, we have separately calculated the detection point distribution of the three methods (CED, CED-OM, CED-CNN) in Weibo-all datasets. From this figure, we find that: 

(1) About $30\%$ microblogs can be detected with less than $10\%$ repost information. More than $40/60\%$ microblogs can be detected with less than $10/20\%$ repost information respectively using CED-OM. It reflects the applicability of our model to handle different rumors. 

(2) With the consideration of original information and repost sequence information using CNN, CED-CNN can detect more rumors than both CED and CED-OM with $10\%$ repost information, that is, more than $60\%$ of samples in Weibo-all dataset. This verifies the advantage of incorporating original microblogs and CNN-forwarding-processing method again.

(3) A peak occurs when using the whole repost sequence to make detections (shown as using $100\%$ repost information). It can be seen that this situation occurs much less in CED-CNN, which is less than $10\%$, indicating that CED-CNN needs less repost information to make credible detection and has higher utilization of forwarding information.

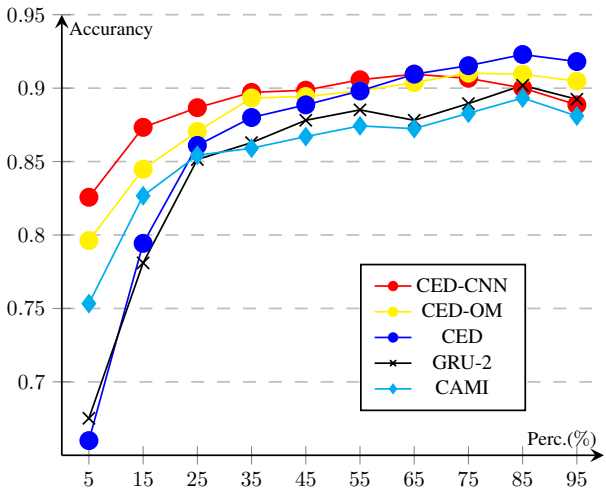
\begin{figure}[!htb]
\centering
		\subfigure{
			\resizebox{0.45\textwidth}{!}{
			\begin{tikzpicture}
			\begin{axis}[
			xlabel={\small Perc.($\%$) },
			ylabel={\small Accurancy},
			xmin= 0, xmax= 100,
			ymin=0.65, ymax=0.95,
			xtick={5, 15, 25, 35, 45, 55, 65, 75, 85, 95},
			ytick={0.55, 0.60, 0.65, 0.70, 0.75, 0.80, 0.85, 0.90, 0.95},
			legend style={at={(0.55,0.10)},anchor=south west},
			ymajorgrids=true,
			grid style=dashed,
			axis lines=center,
            legend image post style={scale=0.5},
            nodes={scale=0.5, transform shape},
			]
            
            \addplot[mark=*,red]
			coordinates {
				(5, 0.8257)(15, 0.8733)(25, 0.8867)(35, 0.8971)(45, 0.8986)
				(55, 0.9057)(65, 0.9095)(75, 0.9067)(85, 0.9001)(95, 0.8886)
			};			
			
			\addplot[mark=*,yellow]
			coordinates {
				(5, 0.7962)(15, 0.8448)(25, 0.8705)(35, 0.8933)(45, 0.8943)
				(55, 0.8981)(65, 0.9038)(75, 0.9102)(85, 0.9095)(95, 0.9048)
			};
            
            \addplot[mark=*,blue]
			coordinates {
				(5, 0.6600)(15, 0.7943)(25, 0.8610)(35, 0.8800)(45, 0.8886)
				(55, 0.8981)(65, 0.9095)(75, 0.9153)(85, 0.9229)(95, 0.9181)
			};	
			
			\addplot[mark=x]
			coordinates {
				(5, 0.6752)(15, 0.7810)(25, 0.8514)(35, 0.8629)(45, 0.8781)
				(55, 0.8852)(65, 0.8781)(75, 0.8895)(85, 0.9019)(95, 0.8924)
			};
			
			\addplot[mark=diamond*,cyan]
			coordinates {
				(5, 0.7533)(15, 0.8267)(25, 0.8543)(35, 0.8591)(45, 0.8670)
				(55, 0.8743)(65, 0.8724)(75, 0.8829)(85, 0.8933)(95, 0.8810)
			};
			
			\legend{CED-CNN, CED-OM, CED, GRU-2, CAMI} 
			
			\end{axis}
			\end{tikzpicture}}
        }
   	\caption{Model comparison using different fixed percentages of repost information in Weibo. (Here, Perc. means percentage of used repost information)}
	\label{fig:comparison}
\end{figure}

In order to verify the performance of various methods with limited repost information, we set different percentages of repost information and compare our models with two best-performed baselines, including GRU-2 and CAMI. Note that we do not only just limit the testing repost information, but also limit the repost information for training. As shown in Fig.~\ref{fig:comparison}, we observe that: CED-OM and CED-CNN achieve better performance when the percentage is less than $55\%$, and CED performs best when the percentage is more than $55\%$. This also proves the necessity of original microblog information and CNN-forwarding-processing method in the early stage. 

It is worth pointing out that our CED method can tell the difference of repost information every microblog needs for credible detection, which is more reasonable than a fixed percentage for all microblogs. So this fixed percentage of repost information experiment is actually not suitable for our method. This also explains the phenomena that CED-OM and CED-CNN performs more poorly as applied fixed percentage arises.

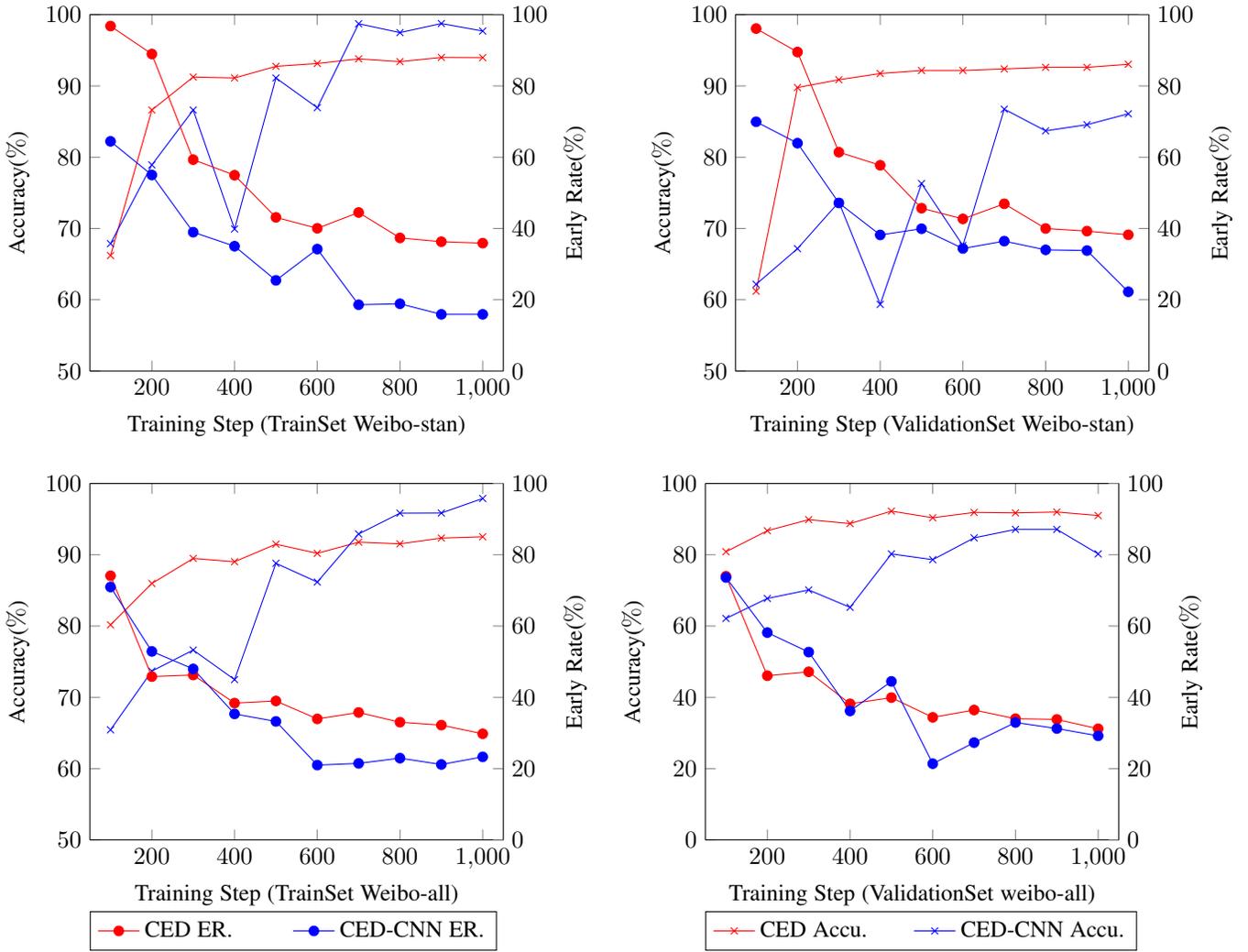
\begin{figure*}
	\centering
	\begin{minipage}{\textwidth}
		\subfigure{
        \begin{tikzpicture}
		\pgfplotsset{
    		scale only axis,
    		scaled x ticks=base 10:0,
    		xmin=50, xmax=1050
		}
		\begin{axis}[
  			axis y line*=left,
  			ymin=50, ymax=100,
  			xlabel = Training Step (TrainSet Weibo-stan),
  			ylabel = Accuracy($\%$),
		]
		\addplot[mark=x,red]
        coordinates{(100,66.20) (200,86.62) (300,91.25) (400,91.12) (500,92.75) (600,93.15) (700,93.80) (800,93.41) (900,94.00) (1000,93.97)
		}; 
        \addplot[mark=x,blue]
        coordinates{(100,67.87) (200,78.92) (300,86.62) (400,69.93) (500,91.12) (600,86.95) (700,98.72) (800,97.48) (900,98.75) (1000,97.70)
		}; 
        \end{axis}
        
		\begin{axis}[
  			axis y line*=right,
  			axis x line=none,
  			ymin=0, ymax=100,
 			ylabel = Early Rate($\%$),
		]
		
        \addplot[mark=*,red]
        coordinates{(100,96.79) (200,88.95) (300,59.30) (400,54.94) (500,43.12) (600,40.05) (700,44.49) (800,37.36) (900,36.29) (1000,35.89)
		}; 
        \addplot[mark=*,blue]
        coordinates{(100,64.44) (200,55.00) (300,38.96) (400,35.03) (500,25.44) (600,34.22) (700,18.60) (800,18.87) (900,15.90) (1000,15.90)
		}; 
		\end{axis}
		\end{tikzpicture}
        }
		\subfigure{
        \begin{tikzpicture}
		\pgfplotsset{
    		scale only axis,
    		scaled x ticks=base 10:0,
    		xmin=50, xmax=1050
		}
		\begin{axis}[
  			axis y line*=left,
  			ymin=50, ymax=100,
  			xlabel = Training Step (ValidationSet Weibo-stan),
  			ylabel = Accuracy($\%$),
		]
		\addplot[mark=x,red]
  		coordinates{(100,61.20) (200,89.78) (300,90.87) (400,91.74) (500,92.17) (600,92.17) (700,92.39) (800,92.61) (900,92.61) (1000,93.04)
		}; 
        \addplot[mark=x,blue]
        coordinates{(100,62.17) (200,67.17) (300,73.70) (400,59.35) (500,76.30) (600,67.61) (700,86.74) (800,83.70) (900,84.56) (1000,86.09)
		}; 
        \end{axis}
        
		\begin{axis}[
  			axis y line*=right,
  			axis x line=none,
  			ymin=0, ymax=100,
 			ylabel = Early Rate($\%$),
		]
		
        \addplot[mark=*,red]
        coordinates{(100,96.09) (200,89.51) (300,61.40) (400,57.73) (500,45.69) (600,42.69) (700,46.92) (800,40.00) (900,39.26) (1000,38.23)
		}; 
        \addplot[mark=*,blue]
        coordinates{(100,69.94) (200,63.93) (300,47.16) (400,38.19) (500,39.93) (600,34.39) (700,36.45) (800,34.01) (900,33.81) (1000,22.23)
		}; 
		\end{axis}
		\end{tikzpicture}
        }
		\subfigure{
        \begin{tikzpicture}
		\pgfplotsset{
    		scale only axis,
    		scaled x ticks=base 10:0,
    		xmin=50, xmax=1050
		}
		\begin{axis}[
  			axis y line*=left,
  			ymin=50, ymax=100,
  			xlabel = Training Step (TrainSet Weibo-all),
  			ylabel = Accuracy($\%$),
		]
		\addplot[mark=x,red]
        coordinates{(100,80.17) (200,85.98) (300,89.49) (400,89.03) (500,91.49) (600,90.21) (700,91.78) (800,91.54) (900,92.35) (1000,92.53)
		}; 
        \label{plot_one}
        \addplot[mark=x,blue]
        coordinates{(100,65.45) (200,73.68) (300,76.63) (400,72.49) (500,88.81) (600,86.18) (700,92.92) (800,95.84) (900,95.86) (1000,97.90)
  		
		}; 
        \label{plot_two}
        \end{axis}
        
		\begin{axis}[
  			axis y line*=right,
  			axis x line=none,
  			ymin=0, ymax=100,
 			ylabel = Early Rate($\%$),
            legend style={
                    at={(0.0,-0.2)},
                    anchor=north west,
                    legend columns=-1,
                    /tikz/every even column/.append style={column sep=0.8cm}
                        },
		]
		
        \addplot[mark=*,red]
        coordinates{(100,74.12) (200,45.82) (300,46.27) (400,38.39) (500,38.99) (600,33.96) (700,35.76) (800,33.03) (900,32.21) (1000,29.78)
		}; 
        \addlegendentry{CED ER.}
        \addplot[mark=*,blue]
        coordinates{(100,70.96) (200,52.88) (300,47.97) (400,35.35) (500,33.25) (600,20.99) (700,21.47) (800,22.94) (900,21.16) (1000,23.30)
        };
        \addlegendentry{CED-CNN ER.}
		\end{axis}
		\end{tikzpicture}
        }
        \subfigure{
        \begin{tikzpicture}
		\pgfplotsset{
    		scale only axis,
    		scaled x ticks=base 10:0,
    		xmin=50, xmax=1050
		}
		\begin{axis}[
  			axis y line*=left,
  			ymin=0, ymax=100,
  			xlabel = Training Step (ValidationSet weibo-all),
  			ylabel = Accuracy($\%$),
		]
		\addplot[mark=x,red]
        coordinates{(100,80.86) (200,86.75) (300,89.86) (400,88.75) (500,92.25) (600,90.38) (700,91.88) (800,91.75) (900,92.00) (1000,91.00)
		}; 
        \addplot[mark=x,blue]
        coordinates{(100,62.13) (200,67.75) (300,70.13) (400,65.25) (500,80.25) (600,78.63) (700,84.75) (800,87.13) (900,87.13) (1000,80.25)
		}; 
        \end{axis}
        
		\begin{axis}[
  			axis y line*=right,
  			axis x line=none,
  			ymin=0, ymax=100,
 			ylabel = Early Rate($\%$),
            legend style={
                    at={(0.0,-0.2)},
                    anchor=north west,
                    legend columns=-1,
                    /tikz/every even column/.append style={column sep=0.6cm}
                        },
		]
		\addlegendimage{/pgfplots/refstyle=plot_one}
        \addlegendentry{CED Accu.}
        \addlegendimage{/pgfplots/refstyle=plot_two}
        \addlegendentry{CED-CNN Accu.}
		
        \addplot[mark=*,red]
        coordinates{(100,73.99) (200,46.07) (300,47.16) (400,38.19) (500,39.93) (600,34.39) (700,36.45) (800,34.01) (900,33.81) (1000,31.19)
		}; 
        \addplot[mark=*,blue]
        coordinates{(100,73.63) (200,58.16) (300,52.69) (400,36.19) (500,44.47) (600,21.38) (700,27.30) (800,32.95) (900,31.25) (1000,29.20)
		}; 
		\end{axis}
		\end{tikzpicture}
        }
	\end{minipage}
	\caption{Early Rate and Accuracy changes in train dataset and validation dataset during training process.}
	\label{fig:train}
\end{figure*}

\begin{figure*}[!htb]
	\centering
	\begin{minipage}{\textwidth}
		\subfigure{\includegraphics[width=0.33\textwidth]{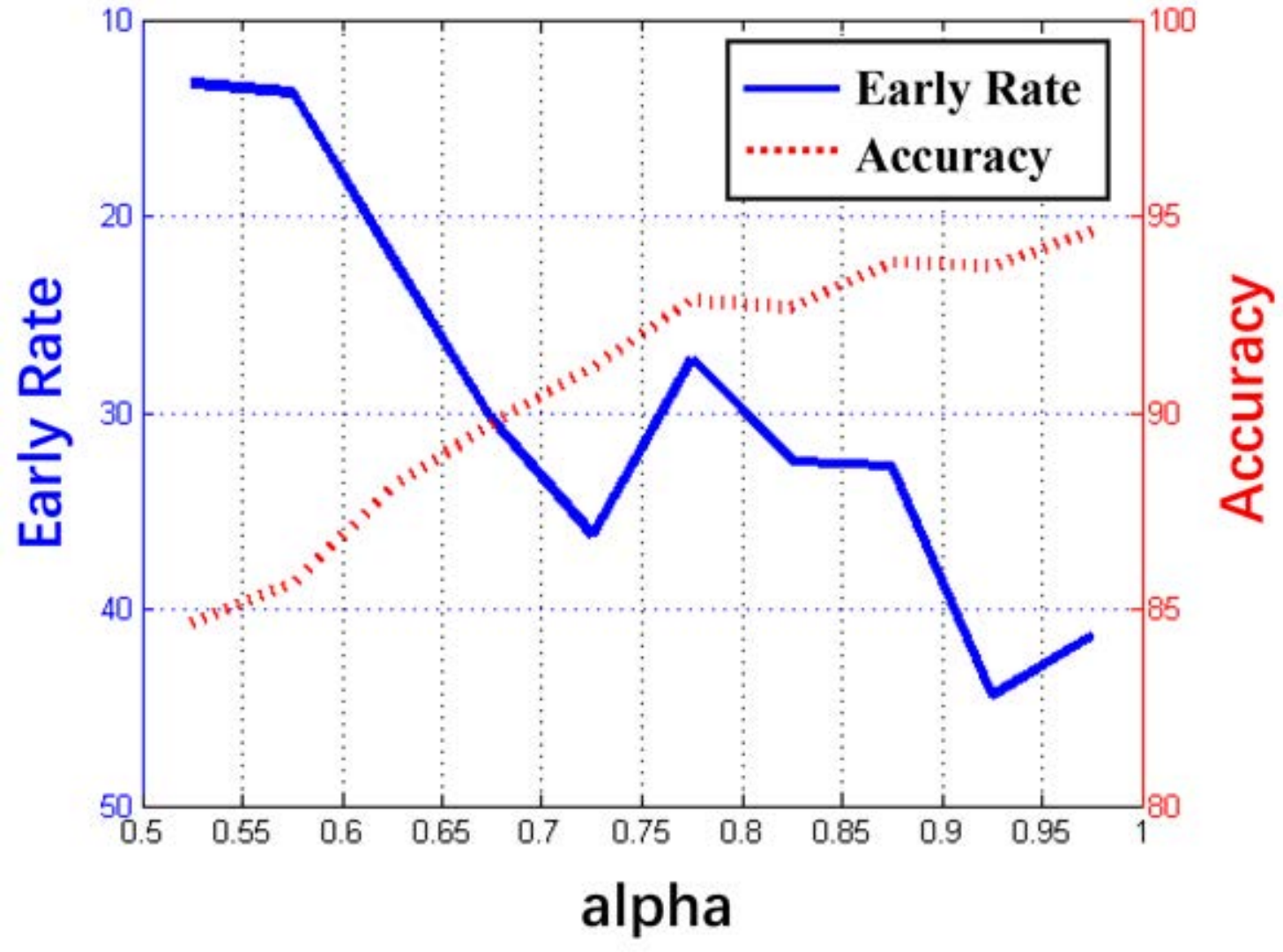}}
		\subfigure{\includegraphics[width=0.33\textwidth]{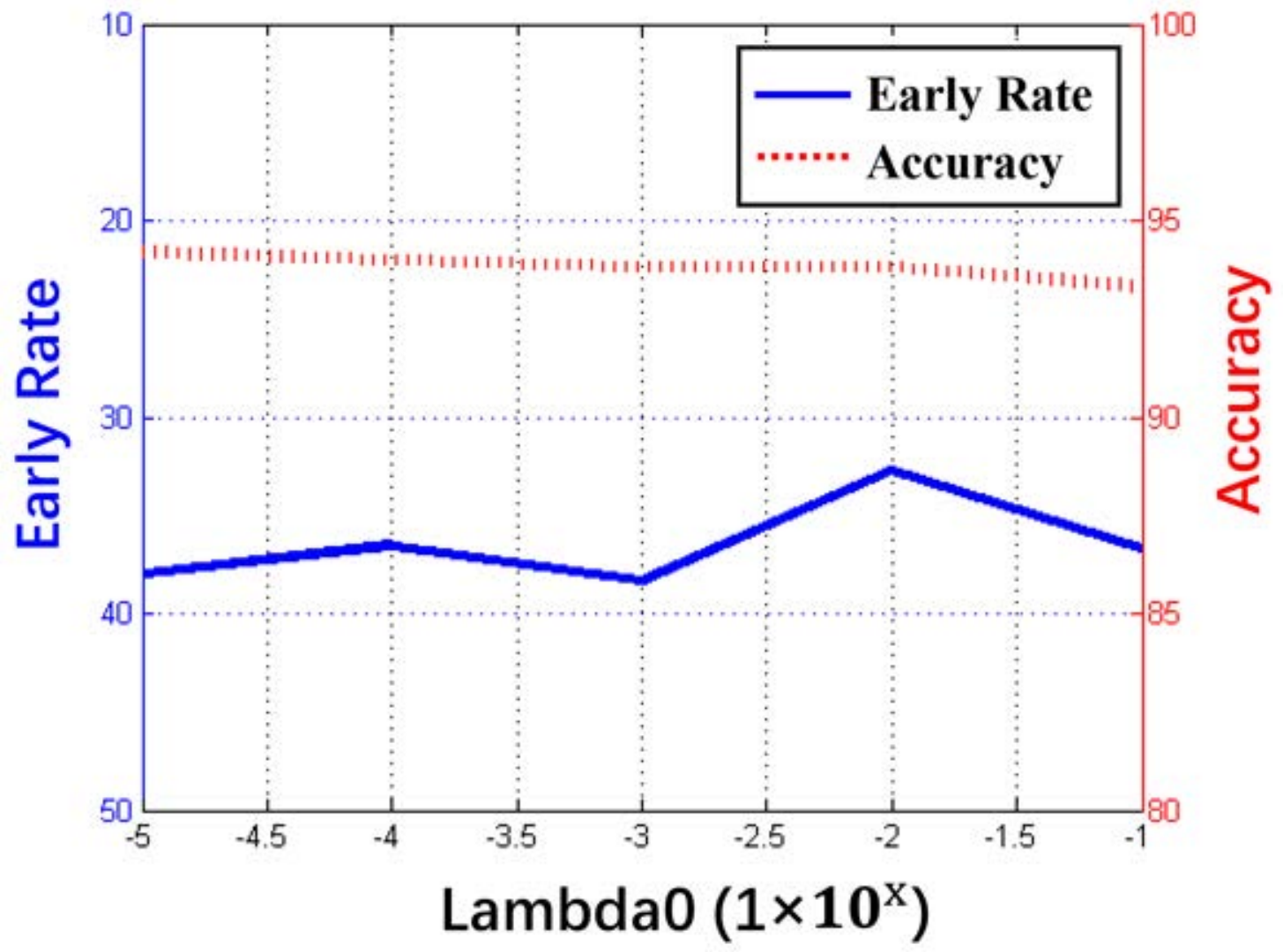}}
		\subfigure{\includegraphics[width=0.33\textwidth]{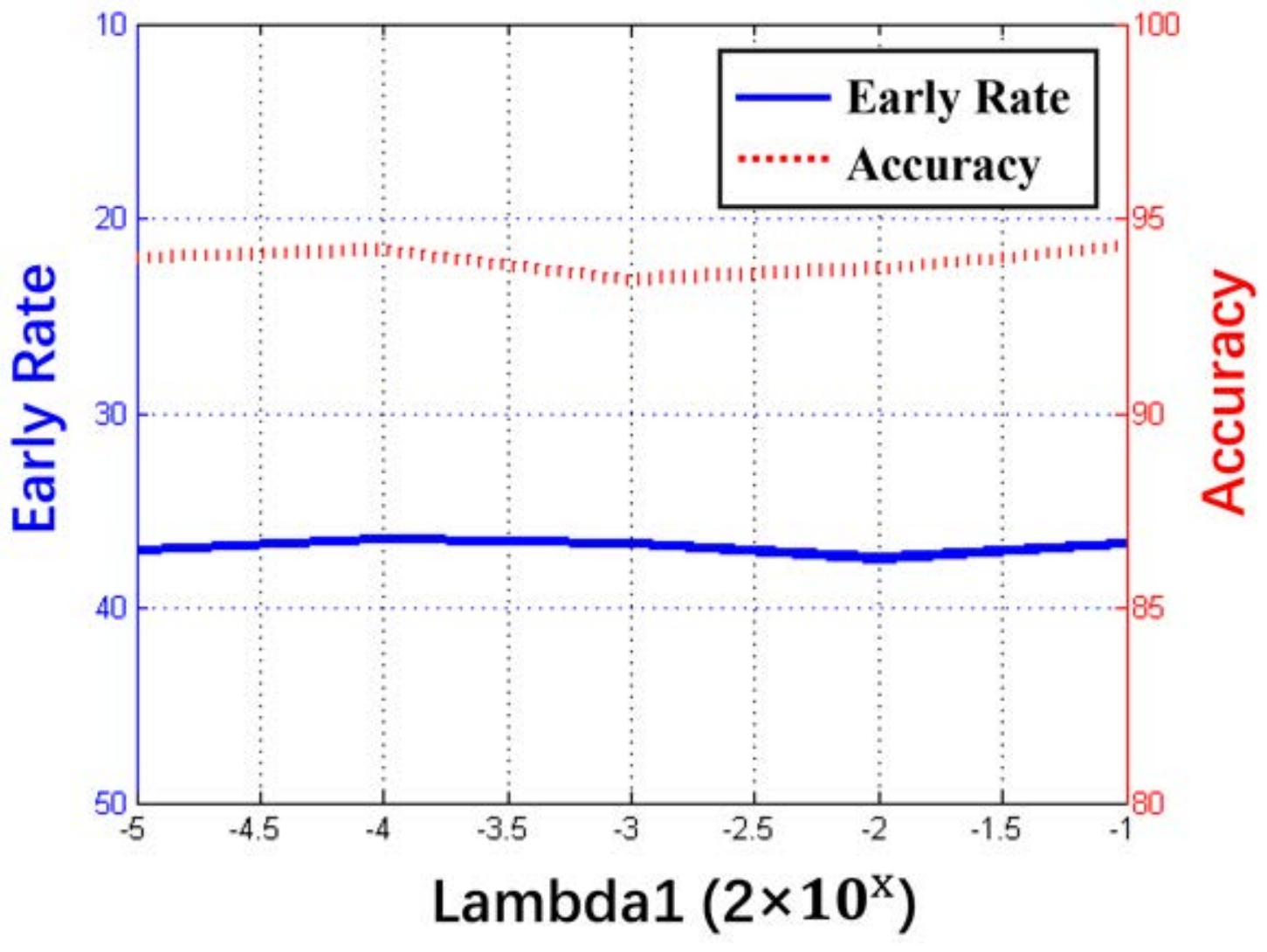}}
        \subfigure{\includegraphics[width=0.33\textwidth]{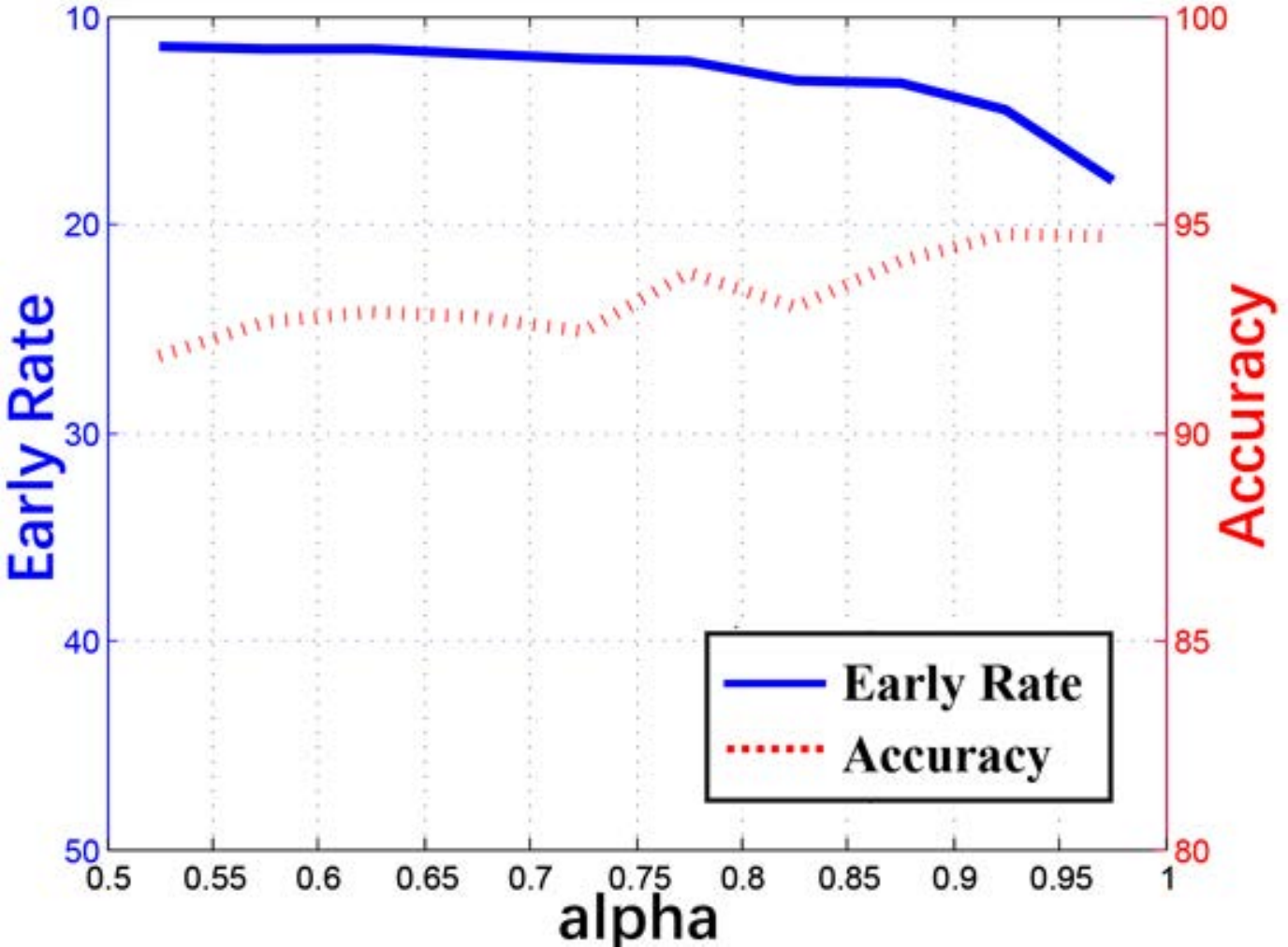}}
		\subfigure{\includegraphics[width=0.33\textwidth]{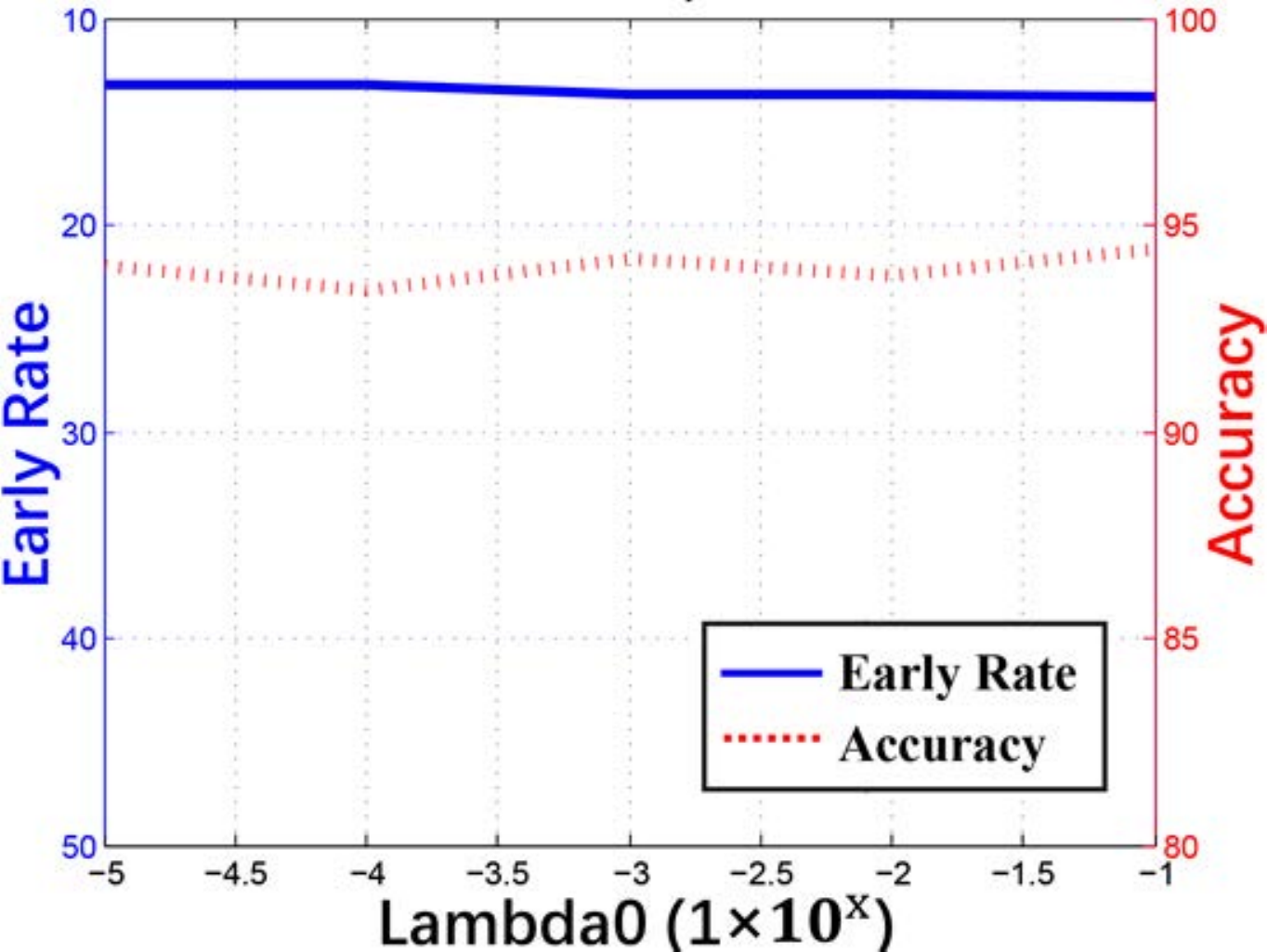}}
		\subfigure{\includegraphics[width=0.33\textwidth]{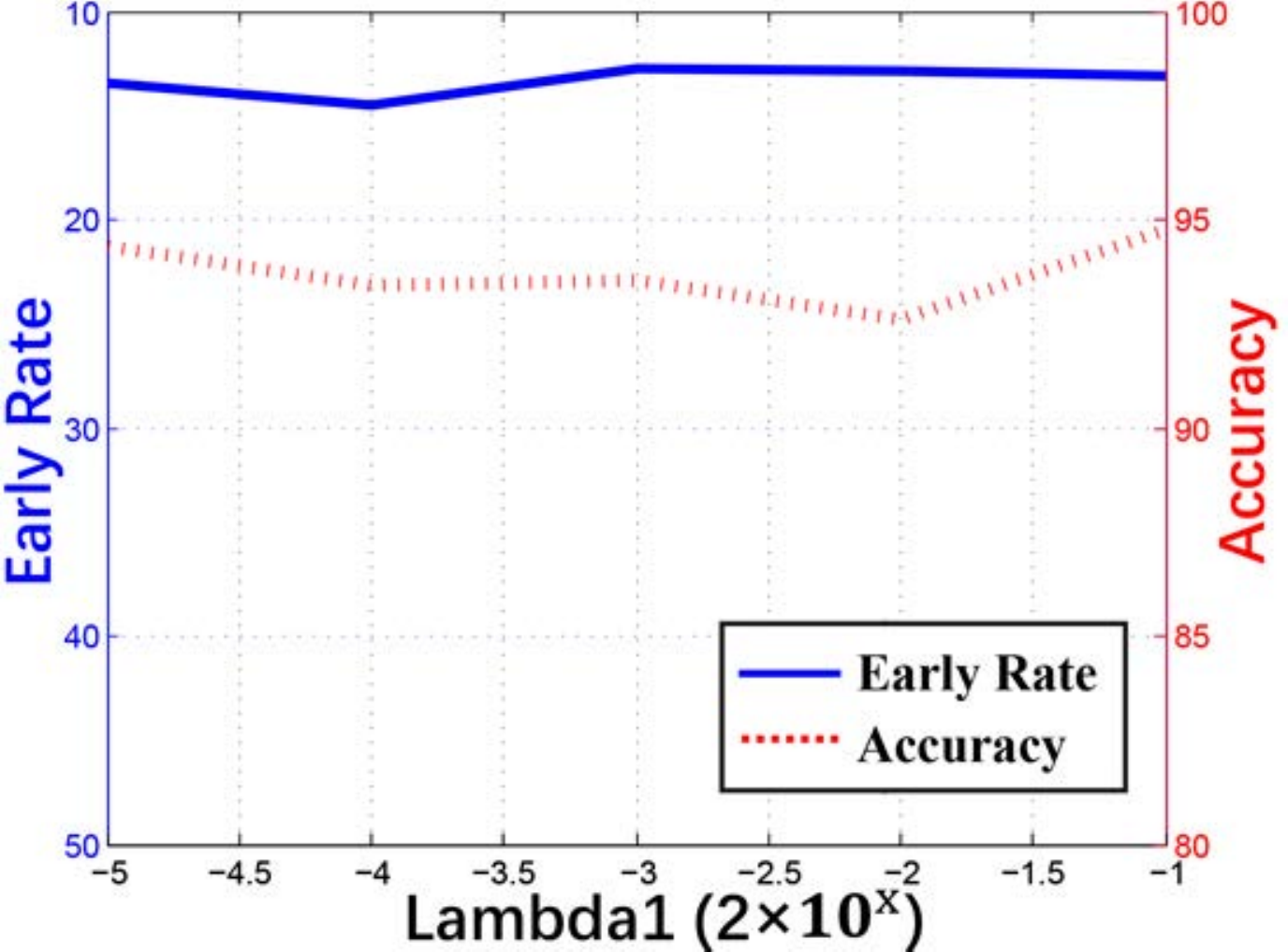}}
	\end{minipage}
	\caption{Parameter Sensitivities of CED (left: $\alpha$; middle: $\lambda_0$; right: $\lambda_1$; top three figures: CED, three figures below: CED-CNN.) }
	\label{fig:parameter}
\end{figure*}

\subsection{Loss Comparison}
In section~\ref{ER}, we explain the idea of ``Credible Detection Point'' applied to objective functions. To verify the role of various parts in the objective function, we perform experiments using different objective functions.

Taking $\mathcal{O}_{time}$ as an example, in order to explore the role of $\mathcal{O}_{time}$ in the model, we design an objective function that lacks $\mathcal{O}_{time}$ :
\begin{equation}
\small
\label{eq:lacktime}
\mathcal{O}_{1} = \mathcal{O}_{pred} + \lambda_0\cdot \mathcal{O}_{diff}.
\end{equation}
We compare this result using $\mathcal{O}_{1}$ with the result of using $\mathcal{O}_{CED}$, which contains the whole objective function parts as Eq.~\ref{eq:model2}. In this way, we can learn about the contribution of $\mathcal{O}_{time}$ to the overall approach.

Similarly, we set up another objective function $\mathcal{O}_{2}$ to verify the effectiveness of $\mathcal{O}_{diff}$:
\begin{equation}
\small
\label{eq:lackdiff}
\mathcal{O}_{2} = \mathcal{O}_{pred} + \lambda_1\cdot \mathcal{O}_{time}.
\end{equation}
We choose CED-CNN method and Weibo-all dataset to finish this experiment. The relevant results are shown in Table~\ref{table:losscompare}.

By comparing the results of $\mathcal{O}_{1}$ and $\mathcal{O}_{2}$ experiments with the overall results $\mathcal{O}_{CED}$, we get the following conclusions:

(1) Compared with the CED-CNN method we proposed, method $\mathcal{O}_{2}$  gets almost the same Accuracy as CED-CNN, but Early Rate decreases by $3\%$, indicating that $\mathcal{O}_{time}$ can enhance model early detection performance without loss of Accuracy.

(2) The Accuracy of $\mathcal{O}_{2}$ method is reduced significantly, and there is also a loss in Early Rate performance, which is compared to the CED-CNN method. This shows that $\mathcal{O}_{diff}$ plays an important role both in promoting Early Rate and Accuracy.

The above experiments effectively validate the specific role of $\mathcal{O}_{time}$ and $\mathcal{O}_{diff}$ in model training, both of which reflect that the purpose of the design is effectively achieved. 

Especially, the designed purpose of $\mathcal{O}_{diff}$ is to minimize the difference value under the upper threshold of rumors, and to minimize the difference exceed the lower threshold of non-rumors after Credible Detection Point(CDP), so that the CDP of the model will advance overall. This explains why the Early Rate performance of $\mathcal{O}_{2}$ is weaker than that of the CED-CNN. Besides, $\mathcal{O}_{diff}$ guarantee the credibility of the detection point. In other words, the prediction probabilities after this detection point should be stable and scale out the threshold, which explains why the CED-CNN's accuracy is significantly higher than that of $\mathcal{O}_{2}$.
	
From this conclusion analysis, we confirm the validity of introducing ``Credible Detection Point'' thought into objective functions.

\subsection{Training process}
In addition to the description of the final effect of the above method, we also conduct a study on the training process of the above two rumor early detection methods: CED and CED-CNN. During the training, every time a batch is provided, the weights are updated in the direction that minimizes the loss, which is defined as a training step. We analyze the training process of the model by observing the performance of the model in training dataset and validation dataset under incremental training step size. Our focus on performance includes Early Rate (ER.) and Accuracy (Accu.). Moreover, We conduct experiments on two Weibo datasets separately, and the results are shown in the Fig~\ref{fig:train}. We can clearly see the following phenomena from the figure. At the same time, we explain the corresponding conclusion:

(1) The accuracy gradually increases and tends to stabilize as the number of training steps increases. At the same time, Early Rate is decreasing as the training process continues.

(2) In the train set of the two datasets, the accuracy in the CED training process reaches a steady state earlier (at about 400 steps), while CED-CNN has great fluctuations in the accuracy performance of the training set. But the accuracy of CED-CNN in the training set can obviously exceed that of CED beginning at about 700 steps. This shows that although CED-CNN may have more ups and downs in training process and need more steps to reach stability, it is more likely to achieve higher accuracy in the training set.

(3) Early Rate performance of CED-CNN is always better than that of CED during entire training process in Weibo-stan dataset, including train set and validation set. But Early Rate performance of CED-CNN is worse than CED in early Weibo-all dataset training (before 300 $\sim$ 400 steps), which shows that the excellent effect of CED-CNN on Early Rate performance can only be achieved after sufficient training, especially in large data sets.

(4) By comparing the performance of CED and CED-CNN in the accuracy of validation dataset, we can clearly see that the stability of CED-CNN in Weibo-all dataset has improved significantly. This indicates the importance of dataset size. The larger the dataset size is, the better the stable selection of the correct optimal model is made in the training process.

\subsection{Parameter Sensitivity.}
There are three critical hyper-parameters in CED, CED-OM, and CED-CNN, i.e., $\alpha$, $\lambda_0$ and $\lambda_1$. To verify the stability of our model, we investigate the parameter sensitivities here. In Fig.~\ref{fig:parameter}, we show the accuracy and early rate results of CED (top three) and CED-CNN (three below) under various parameter settings in Weibo-stan dataset, which is smaller to show parameter sensitivity more intuitively. From this figure, we observe that:

(1) The hyper-parameter $\alpha$ controls the threshold for prediction. From the left two figures in Fig. ~\ref{fig:parameter}, we observe that increasing $\alpha$ can benefit prediction accuracy but lengthen the detection time. Therefore, we need to find a tradeoff between accuracy and early rate according to the real-world scenarios. Based on our experience, when $\alpha=0.975$, our model performs better than baselines and reduces the detection time by about $87\%$.

(2) In CED-CNN, the hyper-parameter $\alpha$ shows consistently good results and excellent stability. This shows that the threshold setting, which directly determines the judgment of early detection point, has an insufficient effect on the performance of the model. CED-CNN perform well in both Early Rate and Accuracy with different $\alpha$, so it is very reliable when balancing Early Rate and Accuracy performance.

(3) The hyper-parameter $\lambda_0$ controls the weight of the difference loss, and the hyper-parameter $\lambda_1$ controls the weight of the prediction time as in Eq. \ref{eq:model2}. From the middle and right figures in Fig.~\ref{fig:parameter}, we find that CED has a stable performance when $\lambda_0$ and $\lambda_1$ range in a large scope (from $10^{-5} $ to $10^{-1}$), for both CED and CED-CNN. It demonstrates the stability and robustness of CED and CED-CNN method.

All the observations above indicates that our model is flexible to different parameter settings and can be easily trained in practice. In addition, we see that the stability of CED-CNN is very excellent and it is much more stable than CED.

\begin{table*}
	\caption{Selected case and their repost information. (CDP indicates ``Credible Detection Point''.)}
	\label{table:case}
	\small
	\centering
	\begin{tabular}{p{0.35\columnwidth}|p{1.3\columnwidth}|p{0.22\columnwidth}}
		\toprule
		\multicolumn{3}{c}{Original Microblog: ``The famous anti-fraud activist Fang Zhouzi announced that he would stop updating in Sina}\\ 
		\multicolumn{3}{c}{Weibo and move to another platform.'' Everyone who reposts this microblog will receive an iPhone.} \\ \midrule
		2012-08-14 13:09:18 & ``\underline{\emph{Good}}! Waiting for the iPhone!'' ``You will \textbf{break} your promise!'' ``\underline{\emph{Reposted}}''&  0.247 \\ \midrule
		2012-08-14 13:15:28 & ``\underline{\emph{Followed}}! Waiting for the phone!'' ``\underline{\emph{Followed}}!'' ``Will I get two phones if I repost twice?'' &  0.265 \\ \midrule
		2012-08-14 13:29:03 & ``\underline{\emph{Okay}}$\sim$ Waiting for your \underline{\emph{gift}}$\sim$'' ``\textbf{Really}?'' ``\textbf{No one} will \textbf{believe} it!'' &  0.563 \\ \midrule
		2012-08-14 13:40:25 & ``@Wenzhou Entertainment News. It's only a \textbf{micro-marketing event}!'' &  \textbf{0.977 (CDP)} \\ \midrule
		2012-08-14 13:45:03 & ``Only \textbf{word games}! Your account should be \textbf{closed} directly!'' &  0.958 \\ \midrule
		2012-08-14 13:51:22 & ``\underline{\emph{Reposted}}! Wonder if it could be realized!‘' &  0.917 \\
		\bottomrule
	\end{tabular}
\end{table*}

\begin{table*}
	\caption{Selected cases and their repost information. (We bold the negative signals and underline the positive ones.) }
	\label{table:case2}
	\centering
	\begin{tabular}{p{0.35\columnwidth}|p{0.55\columnwidth}|p{0.35\columnwidth}|p{0.55\columnwidth}}
		\toprule
		\multicolumn{2}{c|}{\textbf{Correct Case}} &  \multicolumn{2}{|c}{\textbf{Incorrect Case}} \\ \midrule
		\multicolumn{1}{c|}{Repost Time} & \multicolumn{1}{c|}{Content} & \multicolumn{1}{c|}{Repost Time} & \multicolumn{1}{c}{Content} \\ \midrule
		2012.11.17  09:45:30 & It's always a \textbf{trap}! &  2013.12.06    23:40:48  & Save in address book, the latest news! \\ \midrule
		2012.11.17  12:33:43 & Spreading, \textbf{vigilance} &  2013.12.08    14:54:49  & @Changsha City Small Animal Protection Association\\ \midrule
		2012.11.17   12:35:06  & What are the reasons for so many \textbf{liars}? &  2013.12.08    18:24:37  & \underline{Finaly}! \\ \midrule
		2012.11.17   16:56:18  & Near the year off, all kinds of \textbf{trick} are  likely to happen, be sure to be \textbf{careful} &  2013.12.09    00:23:21  & \underline{Great}\\ \midrule
		2012.11.17    19:08:55  & [{surprise}][\textbf{dizzy}]&  2013.12.09    02:10:21  & \underline{Really}? \underline{Really}? \underline{good}!\\ \midrule
		2012.11.18    08:14:16  & \textbf{False} news &  2013.12.09    12:41:05  & \underline{Really} stunned\\ \midrule
		2012.11.21    00:01:36  & Be \textbf{careful}! &  2013.12.09    13:50:44  & [\underline{applaud}][\underline{applaud}] \\ \midrule
		2012.11.24    14:31:26  & \textbf{Not true}, but onlookers&  2013.12.09    19:26:14  & \textbf{Useless} \\ \midrule
		2012.12.12    22:06:21  & \underline{True} or \textbf{false} ~ ~&  2013.12.09    19:59:51  & Remember this phone! \\ \midrule
		2013.01.24    16:13:59  & I received a similar phone call, but fortunately i did \textbf{not believe}.&  2013.12.09    22:06:57  & It is said that the message is \textbf{false} \\ \bottomrule
	\end{tabular}
\end{table*}

\subsection{Case Study and Error Analysis}
To give an intuitive illustration of  CED, we select a representative case from Weibo, which is correctly predicted as a rumor by CED, and show its repost list in Table~\ref{table:case}. In this table, we bold the negative signals and underline the positive ones manually for an intuitive understanding. We also show the prediction probability of each interval in the last column.

From this table, we observe the probabilities of the first several intervals vary a lot due to the conflict information in these intervals. Then, CED makes a correct prediction at the ``Credible Detection Point'', and the rest probabilities are quite consistent. That conforms to the assumption of CED that the prediction curve after ``Credible Prediction Point'' should be stable.

For the incorrectly predicted ones, their reposts are quite unsatisfactory, which leads to difficult to find a credible prediction point and make a credible prediction. We summarize three reasons for errors: (1) Too few reposts; (2) Reposts are quite conflicting; (3) Reposts are not relevant to the original microblogs. 

Actually, (1) and (3) reasons can be solved as time goes by. The important microblog texts that need to be identified as rumors or not will inevitably accumulate enough comments to be used by the CED. Then to demonstrate the conflicting reposts problem of CED, we select two representative cases and show their repost lists in Table~\ref{table:case2}. From this table, we observe that CED can make a correct prediction at an early stage because the rest reposts are quite consistent. That conforms to the assumption of CED that the prediction curve after ``Credible Prediction Point'' should be stable. However, for the incorrect one, its reposts are quite conflicting, which result in a problem that CED cannot find a credible prediction point and make a credible prediction. 

Therefore, we conclude that CED is not good at addressing the conflicting repost sequence well, which will be our future work.

\section{Conclusion and Future Work}
In this paper, we focus on the task of early detection of social media rumors. This task aims to distinguish a rumor as early as possible based on its repost information. While existing works can only make a prediction with the entire or fixed proportions of repost sequence, we assume there exists a ``Credible Detection Point'' for each microblog. Moreover, we propose Credible Early Detection (CED) model to remove the effect of interference repost information and make a credible prediction at a specific credible detection point. Experimental results on real-world datasets demonstrate that our model can significantly reduce the time span for predictions by more than $85\%$, with even better accuracy.

For future works, we can incorporate other important information into early rumor detection, such as publisher's profiles and propagation structure besides the repost information and original microblogs. This additional information is expected to solve the conflict issue in repost sequence.

Another direction is to distinguish the critical signals or reposts from the repost sequence would be helpful for early rumor detection. We will try to utilize attention mechanism to select important reposts and improve the early rumor detection.


\ifCLASSOPTIONcompsoc
  \section*{Acknowledgments}
  This work is supported by the Major Project of the National Social Science Foundation of China under Grant No. $13\&ZD190$. This research is also part of the NExT++ project, supported by the National Research Foundation, Prime Minister’s Office, Singapore under its IRC@Singapore Funding Initiative.
\else
  \section*{Acknowledgment}
  This work is supported by the Major Project of the National Social Science Foundation of China under Grant No. $13\&ZD190$. This research is also part of the NExT++ project, supported by the National Research Foundation, Prime Minister’s Office, Singapore under its IRC@Singapore Funding Initiative.
\fi

\ifCLASSOPTIONcaptionsoff
  \newpage
\fi

\bibliography{reference}

\begin{thebibliography}{10}
\providecommand{\url}[1]{#1}
\csname url@samestyle\endcsname
\providecommand{\newblock}{\relax}
\providecommand{\bibinfo}[2]{#2}
\providecommand{\BIBentrySTDinterwordspacing}{\spaceskip=0pt\relax}
\providecommand{\BIBentryALTinterwordstretchfactor}{4}
\providecommand{\BIBentryALTinterwordspacing}{\spaceskip=\fontdimen2\font plus
\BIBentryALTinterwordstretchfactor\fontdimen3\font minus
  \fontdimen4\font\relax}
\providecommand{\BIBforeignlanguage}[2]{{%
\expandafter\ifx\csname l@#1\endcsname\relax
\typeout{** WARNING: IEEEtran.bst: No hyphenation pattern has been}%
\typeout{** loaded for the language `#1'. Using the pattern for}%
\typeout{** the default language instead.}%
\else
\language=\csname l@#1\endcsname
\fi
#2}}
\providecommand{\BIBdecl}{\relax}
\BIBdecl

\bibitem{allport1947psychology}
G.~W. Allport and L.~Postman, ``The psychology of rumor.'' 1947.

\bibitem{kapferer1987rumeurs}
J.-N. Kapferer, \emph{Rumeurs: le plus vieux m{\'e}dia du monde}.\hskip 1em
  plus 0.5em minus 0.4em\relax Editions du seuil, 1987.

\bibitem{peterson1951rumor}
W.~A. Peterson and N.~P. Gist, ``Rumor and public opinion,'' \emph{American
  Journal of Sociology}, vol.~57, no.~2, pp. 159--167, 1951.

\bibitem{liu2015rumor}
Z.~Liu, L.~Zhang, C.~Tu, and M.~Sun, ``Statistical and semantic analysis of
  rumors in chinese social media,'' \emph{Science China: Informationis}, 2015.

\bibitem{castillo2011information}
C.~Castillo, M.~Mendoza, and B.~Poblete, ``Information credibility on
  twitter,'' in \emph{Proceedings of WWW}, 2011, pp. 675--684.

\bibitem{yang2012automatic}
F.~Yang, Y.~Liu, X.~Yu, and M.~Yang, ``Automatic detection of rumor on sina
  weibo,'' in \emph{Proceedings of KDD}, 2012, p.~13.

\bibitem{kwon2013prominent}
S.~Kwon, M.~Cha, K.~Jung, W.~Chen, and Y.~Wang, ``Prominent features of rumor
  propagation in online social media,'' in \emph{Proceedings of ICDM}, 2013,
  pp. 1103--1108.

\bibitem{liu2015real}
X.~Liu, A.~Nourbakhsh, Q.~Li, R.~Fang, and S.~Shah, ``Real-time rumor debunking
  on twitter,'' in \emph{Proceedings of CIKM}, 2015, pp. 1867--1870.

\bibitem{ma2015detect}
J.~Ma, W.~Gao, Z.~Wei, Y.~Lu, and K.-F. Wong, ``Detect rumors using time series
  of social context information on microblogging websites,'' in
  \emph{Proceedings of CIKM}, 2015, pp. 1751--1754.

\bibitem{wu2015false}
K.~Wu, S.~Yang, and K.~Q. Zhu, ``False rumors detection on sina weibo by
  propagation structures,'' in \emph{Proceedings of ICDE}.\hskip 1em plus 0.5em
  minus 0.4em\relax IEEE, 2015, pp. 651--662.

\bibitem{ma2016detecting}
J.~Ma, W.~Gao, P.~Mitra, S.~Kwon, B.~J. Jansen, K.-F. Wong, and M.~Cha,
  ``Detecting rumors from microblogs with recurrent neural networks.'' in
  \emph{Proceedings of IJCAI}, 2016, pp. 3818--3824.

\bibitem{yu2017convolutional}
F.~Yu, Q.~Liu, S.~Wu, L.~Wang, and T.~Tan, ``A convolutional approach for
  misinformation identification,'' in \emph{Proceedings of IJCAI}, 2017.

\bibitem{le2014distributed}
Q.~Le and T.~Mikolov, ``Distributed representations of sentences and
  documents,'' in \emph{International Conference on Machine Learning}, 2014,
  pp. 1188--1196.

\bibitem{kumar2018false}
S.~Kumar and N.~Shah, ``False information on web and social media: A survey,''
  \emph{arXiv preprint arXiv:1804.08559}, 2018.

\bibitem{bhatt2018combining}
G.~Bhatt, A.~Sharma, S.~Sharma, A.~Nagpal, B.~Raman, and A.~Mittal, ``Combining
  neural, statistical and external features for fake news stance
  identification,'' in \emph{Companion of the The Web Conference 2018 on The
  Web Conference 2018}.\hskip 1em plus 0.5em minus 0.4em\relax International
  World Wide Web Conferences Steering Committee, 2018, pp. 1353--1357.

\bibitem{ruchansky2017csi}
N.~Ruchansky, S.~Seo, and Y.~Liu, ``Csi: A hybrid deep model for fake news
  detection,'' in \emph{Proceedings of the 2017 ACM on Conference on
  Information and Knowledge Management}.\hskip 1em plus 0.5em minus 0.4em\relax
  ACM, 2017, pp. 797--806.

\bibitem{vosoughi2018spread}
S.~Vosoughi, D.~Roy, and S.~Aral, ``The spread of true and false news online,''
  \emph{Science}, vol. 359, no. 6380, pp. 1146--1151, 2018.

\bibitem{shao2016hoaxy}
C.~Shao, G.~L. Ciampaglia, A.~Flammini, and F.~Menczer, ``Hoaxy: A platform for
  tracking online misinformation,'' in \emph{Proceedings of the 25th
  international conference companion on world wide web}.\hskip 1em plus 0.5em
  minus 0.4em\relax International World Wide Web Conferences Steering
  Committee, 2016, pp. 745--750.

\bibitem{ma2017detect}
J.~Ma, W.~Gao, and K.-F. Wong, ``Detect rumors in microblog posts using
  propagation structure via kernel learning,'' in \emph{Proceedings of the 55th
  Annual Meeting of the Association for Computational Linguistics (Volume 1:
  Long Papers)}, vol.~1, 2017, pp. 708--717.

\bibitem{zhao2015enquiring}
Z.~Zhao, P.~Resnick, and Q.~Mei, ``Enquiring minds: Early detection of rumors
  in social media from enquiry posts,'' in \emph{Proceedings of WWW}, 2015, pp.
  1395--1405.

\bibitem{nguyen2017early}
T.~N. Nguyen, C.~Li, and C.~Nieder{\'e}e, ``On early-stage debunking rumors on
  twitter: Leveraging the wisdom of weak learners,'' in \emph{Proceedings of
  ICSI}, 2017, pp. 141--158.

\bibitem{wu2017gleaning}
L.~Wu, J.~Li, X.~Hu, and H.~Liu, ``Gleaning wisdom from the past: Early
  detection of emerging rumors in social media,'' in \emph{Proceedings of
  ICDM}, 2017, pp. 99--107.

\bibitem{salton1988term}
G.~Salton and C.~Buckley, ``Term-weighting approaches in automatic text
  retrieval,'' \emph{Information processing \& management}, vol.~24, no.~5, pp.
  513--523, 1988.

\bibitem{kalchbrenner2014convolutional}
N.~Kalchbrenner, E.~Grefenstette, and P.~Blunsom, ``A convolutional neural
  network for modelling sentences,'' \emph{arXiv preprint arXiv:1404.2188},
  2014.

\bibitem{liu2015convolutional}
Q.~Liu, F.~Yu, S.~Wu, and L.~Wang, ``A convolutional click prediction model,''
  in \emph{Proceedings of the 24th ACM International on Conference on
  Information and Knowledge Management}.\hskip 1em plus 0.5em minus 0.4em\relax
  ACM, 2015, pp. 1743--1746.

\bibitem{kim2014convolutional}
Y.~Kim, ``Convolutional neural networks for sentence classification,''
  \emph{arXiv preprint arXiv:1408.5882}, 2014.

\bibitem{tamar2016value}
A.~Tamar, Y.~Wu, G.~Thomas, S.~Levine, and P.~Abbeel, ``Value iteration
  networks,'' in \emph{Advances in Neural Information Processing Systems},
  2016, pp. 2154--2162.

\bibitem{glorot2011deep}
X.~Glorot, A.~Bordes, and Y.~Bengio, ``Deep sparse rectifier neural networks,''
  in \emph{Proceedings of the Fourteenth International Conference on Artificial
  Intelligence and Statistics}, 2011, pp. 315--323.

\bibitem{cho2014properties}
K.~Cho, B.~Van~Merri{\"e}nboer, D.~Bahdanau, and Y.~Bengio, ``On the properties
  of neural machine translation: Encoder-decoder approaches,'' in
  \emph{Proceedings of SSST-8}, 2014.

\bibitem{dos2016attentive}
C.~N. dos Santos, M.~Tan, B.~Xiang, and B.~Zhou, ``Attentive pooling
  networks,'' \emph{CoRR, abs/1602.03609}, vol.~2, no.~3, p.~4, 2016.

\bibitem{tu2017cane}
C.~Tu, H.~Liu, Z.~Liu, and M.~Sun, ``Cane: Context-aware network embedding for
  relation modeling,'' in \emph{Proceedings of ACL}, 2017.

\bibitem{kingma2014adam}
D.~Kingma and J.~Ba, ``Adam: A method for stochastic optimization,'' in
  \emph{Proceedings of ICLR}, 2015.

\bibitem{suykens1999least}
J.~A. Suykens and J.~Vandewalle, ``Least squares support vector machine
  classifiers,'' \emph{Neural processing letters}, vol.~9, no.~3, pp. 293--300,
  1999.

\end{thebibliography}
\bibliographystyle{IEEEtran}

\begin{IEEEbiography}[{\includegraphics[width=1in, height=1.25in, clip, keepaspectratio]{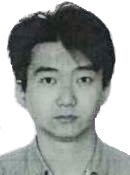}}]{Changhe Song}
is an undergraduate student of the Department of Electronic Engineering, Tsinghua University, Beijing 100084, China. His research interests are natural language processing and social computation.
\end{IEEEbiography}

\begin{IEEEbiography}[{\includegraphics[width=1in, height=1.25in, clip, keepaspectratio]{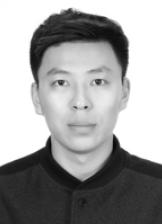}}]{Cunchao Tu}
is a PhD student of the Department of Computer Science and Technology, Tsinghua University. He got his BEng degree in 2013 from the Department of Computer Science and Technology, Tsinghua University. His research interests are natural language processing and social computing. He has published several papers in international conferences and journals including ACL, IJCAI, EMNLP, COLING, and ACM TIST.
\end{IEEEbiography}

\begin{IEEEbiography}[{\includegraphics[width=1in, height=1.25in, clip, keepaspectratio]{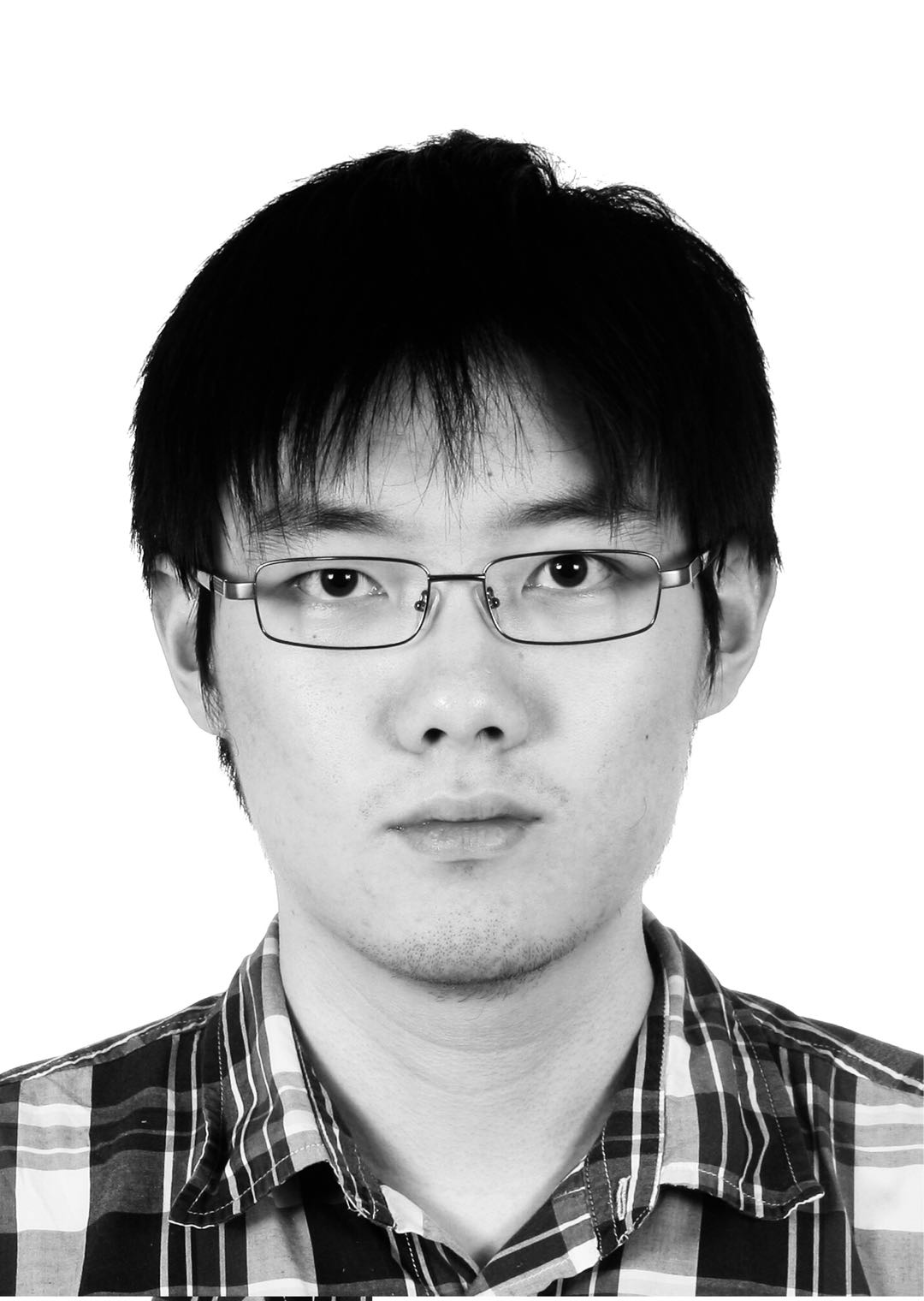}}]{Cheng Yang}
is a 4-th year PhD student of the Department of Computer Science and Technology, Tsinghua University. He got his B.E. degree from Tsinghua University in 2014. His research interests include natural language processing and network representation learning. He has published several top-level papers in international journals and conferences including ACM TOIS, IJCAI and AAAI. 
\end{IEEEbiography}

\begin{IEEEbiography}[{\includegraphics[width=1in, height=1.25in, clip, keepaspectratio]{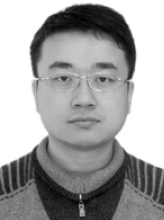}}]{Zhiyuan Liu}
is an associate professor of the Department of Computer Science and Technology, Tsinghua University. He got his BEng degree in 2006 and his Ph.D. in 2011 from the Department of Computer Science and Technology, Tsinghua University. His research interests are natural language processing and social computation. He has published over 40 papers in international journals and conferences including ACM Transactions, IJCAI, AAAI, ACL and EMNLP.
\end{IEEEbiography}

\begin{IEEEbiography}[{\includegraphics[width=1in, height=1.25in, clip, keepaspectratio]{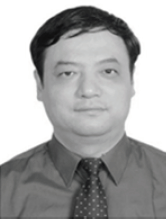}}]{Maosong Sun}
is a professor of the Department of Computer Science and Technology, Tsinghua University. He got his BEng degree in 1986 and MEng degree in 1988 from Department of Computer Science and Technology, Tsinghua University, and got his Ph.D. degree in 2004 from Department of Chinese, Translation, and Linguistics, City University of Hong Kong. His research interests include natural language processing, Chinese computing, Web intelligence, and computational social sciences. He has published over 150 papers in academic journals and international conferences in the above fields. He serves as a vice president of the Chinese Information Processing Society, the council member of China Computer Federation, the director of Massive Online Education Research Center of Tsinghua University, and the Editor-in-Chief of the Journal of Chinese Information Processing.
\end{IEEEbiography}

\setkeys{Gin}{draft=false}
\end{document}